\documentclass[journal, 10pt]{IEEEtran}
\IEEEoverridecommandlockouts

\usepackage{amsmath,amssymb,amsfonts}
\usepackage{graphicx}
\usepackage{textcomp}
\usepackage{xcolor}
\usepackage[style=ieee]{biblatex}

\usepackage{amsmath,amssymb}  
\usepackage{amsthm}  
\usepackage{algorithm}
\usepackage{algpseudocode}
\usepackage{subcaption}
\usepackage{booktabs}
\usepackage{makecell}

\newtheorem{proposition}{Proposition}  
\def\BibTeX{{\rm B\kern-.05em{\sc i\kern-.025em b}\kern-.08em
    T\kern-.1667em\lower.7ex\hbox{E}\kern-.125emX}}
\begin{document}
\title{Robust Multi-modal Task-oriented Communications with Redundancy-aware Representations\\

}

\author{Jingwen Fu,~\IEEEmembership{Student Member,~IEEE,}
Ming Xiao,~\IEEEmembership{Senior Member,~IEEE,}
Zhonghao Lyu,~\IEEEmembership{Member,~IEEE,}
Mikael Skoglund,~\IEEEmembership{Fellow,~IEEE,}
Celimuge Wu,~\IEEEmembership{Senior Member,~IEEE}

\thanks{Jingwen Fu, Ming Xiao, Zhonghao Lyu, and Mikael Skoglund are with the School of Electrical Engineering and Computer Science (EECS), KTH Royal Institute of Technology, 11428 Stockholm, Sweden. (Corresponding author: Ming Xiao.) Email: \{jingwenf, mingx, lzhon, skoglund\}@kth.se.} 
\thanks{Celimuge Wu is with the Department of Computer and Network Engineering, The University of Electro-Communications, 182-8585 Tokyo, Japan. Email: celimuge@uec.ac.jp.}

}

\maketitle

\begin{abstract}
Semantic communications for multi-modal data can transmit task-relevant information efficiently over noisy and bandwidth-limited channels. However, a key challenge is to simultaneously compress inter-modal redundancy and improve semantic reliability under channel distortion. To address the challenge, we propose a robust and efficient multi-modal task-oriented communication framework that integrates a two-stage variational information bottleneck (VIB) with mutual information (MI) redundancy minimization. In the first stage, we apply uni-modal VIB to compress each modality separately, i.e., text, audio, and video, while preserving task-specific features. To enhance efficiency, an MI minimization module with adversarial training is then used to suppress cross-modal dependencies and to promote complementarity rather than redundancy. In the second stage, a multi-modal VIB is further used to compress the fused representation and to enhance robustness against channel distortion. Experimental results on multi-modal emotion recognition tasks demonstrate that the proposed framework significantly outperforms existing baselines in accuracy and reliability, particularly under low signal-to-noise ratio regimes. Our work provides a principled framework that jointly optimizes modality-specific compression, inter-modal redundancy, and communication reliability.
\end{abstract}


\section{Introduction}

The rapid advance of deep learning has precipitated a paradigm shift in the design of communication systems~\cite{deepsc}. Conventional communication systems have been primarily concerned with reliable symbol transmission, i.e., a problem identified as the first level of communication by Shannon and Weaver ~\cite{shannon}. However, the increasing prevalence of intelligent applications is driving a shift from this classical paradigm towards high-level communication objectives~\cite{vl-vfe}. The first of these, semantic communication (SemCom), is used to accurately convey the intended \textit{meaning} of the source data, rather than to ensure mere bit-level fidelity~\cite{deepsc}. A further evolution is task-oriented communication (TOC), where the objective is to transmit the minimal information required for a recipient to successfully execute a specific task. For many emerging applications, such as autonomous driving and remote healthcare, information from a single modality is often insufficient to characterize complex real-world events. Consequently, multi-modal TOC or SemCom has emerged as a critical and rapidly advancing area of research~\cite{guo2024distributed}.

To achieve efficient multi-modal TOC/SemCom, a few frameworks have been proposed. For instance, \cite{vqa} has integrated visual and textual data for visual question-answering (VQA) communication systems. Similarly,  unified frameworks to transmit shared modalities like images, text, and audio have been developed in~\cite{multi_modal_task}. In these systems, the information bottleneck (IB) is shown to be a powerful tool for balancing the trade-off between information compression and task performance~\cite{vl-vfe}. Moreover, methods based on variational information bottleneck (VIB)~\cite{vl-vfe} and robust information bottleneck (RIB)~\cite{rib} have further provided a tractable upper bound for high-dimensional data compression, and coded redundancy reduction to improve transmission efficiency, respectively.
These pioneering works have laid a solid foundation for extracting and transmitting task-relevant semantic information.

Despite significant progress, existing multi-modal TOC frameworks face two critical challenges. First, although current models effectively compress intra-modality information, they often neglect the redundancy across different modalities. For example, in a video segment, visual modality and audio modality may convey the same positive sentiment. Transmitting both  may waste channel bandwidth. Second, the fused multi-modal representations are typically vulnerable to channel impairments, such as noise and fading, which can significantly compromise semantic reliability and degrade task performance. These issues motivate a key research question: How can a multi-modal TOC framework be designed to effectively mitigate inter-modal redundancy and simultaneously enhance the robustness of the fused semantic features against adverse channel conditions such as noise and fading?

To tackle these challenges, we propose a novel framework that integrates a two-stage VIB with mutual information (MI) redundancy minimization. The proposed method follows a hierarchical processing architecture. First, we apply a dedicated uni-modal VIB for each modality, i.e., text, audio, and video, as the first-stage bottleneck to extract modality-specific and task-relevant features. Then, after feature fusion, we introduce a novel cross-modal MI discriminator module, which adversarially suppresses inter-modal dependencies through discriminator training. The module encourages complementary, rather than redundant, information across modalities. Finally, the fused multi-modal representation is further compressed by the second-stage multi-modal VIB to enhance robustness against channel distortion. This end-to-end design, which explicitly incorporates the wireless channel into training, enhances robustness and preserves semantic integrity under channel distortion.
The main contributions are summarized as follows.
\begin{itemize}
    \item We propose a novel two-stage VIB architecture for a multi-modal TOC system, which explicitly considers both modality-specific and fused multi-modal representations via VIB rate-distortion regularization. The proposed hierarchical architecture enables more efficient compression of task-relevant semantics and simultaneously improves robustness against channel distortion.
    \item Our work introduces a cross-modal redundancy reduction module that minimizes MI among modality pairs. Specifically, we formulate pairwise MI among modality representations and further derive and prove a bounded variational lower bound on MI, ensuring both theoretical soundness and training stability. To stabilize the adversarial learning process, we incorporate a gradient reversal layer (GRL) that enables efficient and unified end-to-end training, where encoders are optimized to suppress redundant information while discriminators tighten the MI estimates. This mechanism effectively reduces redundancy among modalities, yielding complementary and compact multi-modal representations.
    \item Extensive experiments on multi-modal tasks show that the proposed framework effectively reduces redundancy, improves transmission robustness under varying channel conditions, and consistently outperforms benchmark in both accuracy and robustness.
\end{itemize}

The remainder of this paper is organized as follows. Section \ref{section:literature} reviews related work. Section \ref{sec:model} presents our system model and problem formulation. Sections \ref{sec:VIB} and \ref{sec:MI} detail the proposed scheme. Section \ref{sec:experiment} presents the experimental setups and results. Finally, Section \ref{sec:conclusion} concludes the paper.

Notations: In what follows, upper-case letters (e.g. $X$) stand for random variables, and lower-case letters (e.g. $x$) are their realizations.  The Kullback-Leibler (KL) divergence between two probability distributions \( p(x) \) and \( q(x) \) is denoted as \( D_{\text{KL}}(p \| q) \). The Jensen-Shannon (JS) divergence is denoted as \( D_{\text{JS}}(p \| q) \). The statistical expectation of \( X \) is denoted as \( \mathbb{E}(X) \).

\section{Related Works}
\label{section:literature}
\subsection{Semantic and Task-oriented Communications}
The classic communication theory of Shannon defines three levels of transmission targets, i.e., technical, semantic, and effectiveness \cite{shannon}. 
Traditional systems focused on the technical level (bit-accuracy). Recent advances in deep learning have prompted a paradigm shift towards semantic transmission~\cite{deepsc}. 
SemCom aims to transmit the meaning of data rather than the raw data, as exemplified by the DeepSC system that uses neural encoders to extract textual semantics with improved efficiency~\cite{deepsc}. 
Building on the principle, TOC further refines the goal to complete a specific task, pruning task-irrelevant information as redundancy~\cite{fu2025computation,10388062}. 

As intelligent applications increasingly integrate heterogeneous data sources, multi-modal SemCom/TOC has become an important research area. 
Researchers have developed multi-modal SemCom systems for diverse scenarios. For example, the authors in  \cite{vqa} have combined images and text in a VQA SemCom system for multi-user settings. Reference \cite{multi_modal_task} proposed a unified end-to-end framework to jointly handle image, text, and audio. Moreover, the authors in \cite{guo2024distributed} have designed distributed architectures for multi-modal semantic relays at network edges. 
These efforts underscore the growing need to efficiently handle heterogeneous data streams in communication networks. More recently, researchers have leveraged artificial intelligence (AI) models to further advance multi-modal SemCom. For example, \cite{jiang2024large} has integrated large pre-trained multi-modal and language models to align textual and visual features, and applied generative adversarial networks (GAN) for channel state estimation. Such large AI model-based approaches provide new opportunities to address challenges of data heterogeneity and lossy channels in multi-modal transmission. In summary, SemCom/TOC has evolved from single-modal systems to sophisticated multi-modal frameworks. However, fully exploiting cross-modal synergies and avoiding redundancy still remains an open challenge.

\subsection{Information Bottleneck in Communication Systems}
The IB principle provides a theoretical framework for extracting the maximally compressed representation of a source variable that remains most informative to achieve a target task~\cite{tishby1999information}. IB learns a bottleneck representation that minimizes the MI  while maximizing task-relevant information. Although theoretically powerful, direct optimization of these MI terms is generally intractable for high-dimensional data. To address the problem, \cite{dvib} has introduced the VIB, a tractable deep learning-based approximation of IB. The VIB framework has been applied to wireless communications to design end-to-end learning-based TOC systems \cite{vl-vfe}.
Since then, VIB has been extended to address diverse challenges and motivate various applications. For instance, VIB has enabled efficient representation learning for multi-device cooperative inference, where multiple agents collaborate on a task \cite{multi_ib}. To address the channel impairments, RIB incorporates coded redundancy and digital modulation to improve reliability in noisy channels \cite{rib}. Despite advances, existing works primarily focus on single modality compression or fused representation. Explicitly addressing cross-modal redundancy through the IB principle remains an under-explored but critical research direction.

\subsection{Multi-modal Fusion and Redundancy Reduction}

\begin{figure*}[t]
\centering
\includegraphics[width=0.85\linewidth]{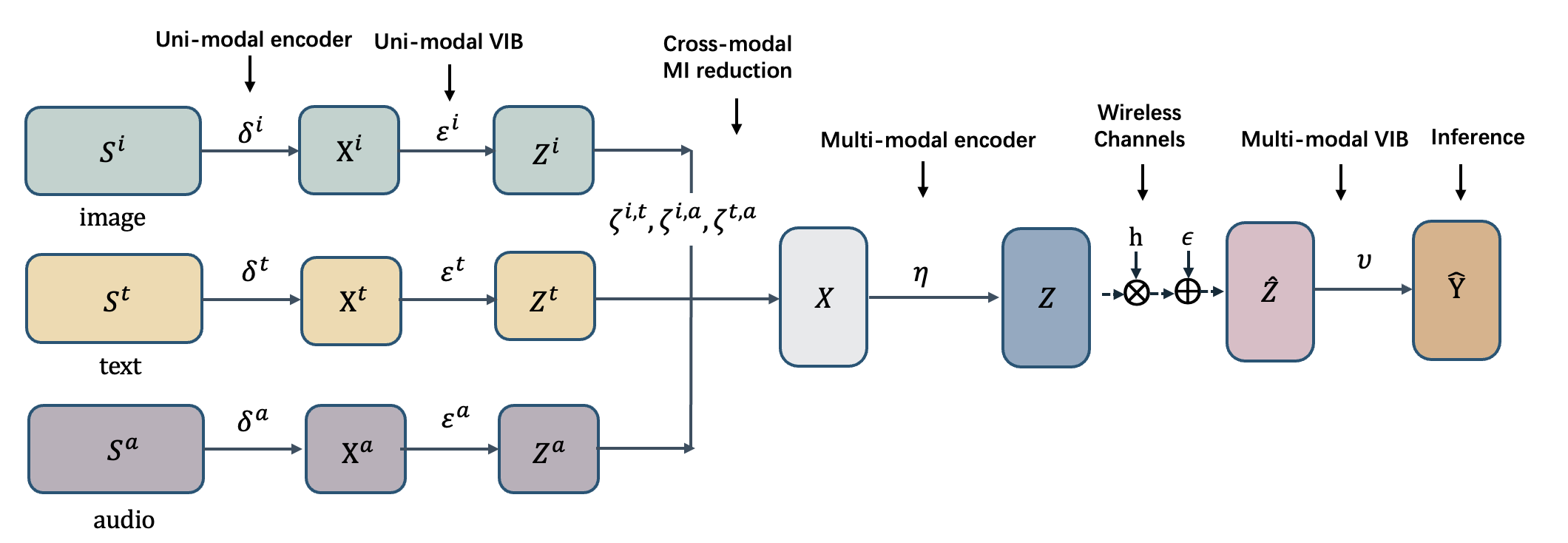}
\caption{The proposed framework for multi-modal TOC.}
\label{frame}
\end{figure*}

A key challenge in multi-modal systems lies in effectively fusing information from different modalities. Common strategies, including early, late, and hybrid fusion, combine modality-specific features into a joint representation that is intended to be more informative than any single modality \cite{jiao2024survey}. However, such straight fusion often introduces redundancy, as different modalities often capture overlapping information about the same task. For example, the visual appearance and the sound of rain both indicate the same weather condition. Transmitting such redundant information not only wastes bandwidth but also reduces system robustness.

To address the challenge, we focus on explicitly minimizing the MI between the representations of different modalities. Although estimating and minimizing MI in high-dimensional spaces is non-trivial, recent advances in variational methods have made it tractable. In particular, JS divergence-based methods, such as those used in f-GAN \cite{nowozin2016f} and Deep InfoMax~\cite{hjelm2019learning}, have provided stable and scalable MI estimators. Furthermore, adversarial training techniques, particularly   GRL~\cite{ganin2016domain}, have provided elegant mechanisms to realize the min–max optimization for adversarial training. To the best of our knowledge, our work is the first to integrate MI redundancy minimization methods into a two-stage VIB framework, especially for robust and efficient multi-modal TOC.

\section{System Model}
\label{sec:model}

The proposed multi-modal TOC framework is depicted in Fig.~\ref{frame}. Specifically, we consider three input modalities: image $i$, text $t$, and audio $a$. For each modality \(m \in \{ i, t, a \}\), we denote the input as \((S^{m}, Y)\), where \(S^{m}\) denotes the raw input data and \(Y\) is its corresponding label. The objective is to learn a mapping from \(S^{m}\)  to a task-specific label $Y$. The end-to-end architecture consists of a transmitter, wireless channels, and a receiver, which are detailed below.

\subsection{Transmitter Architecture}
The transmitter is responsible for extracting a compact and task-relevant latent representation $Z$ from the multi-modal inputs by three stages:

\subsubsection{Uni-modal Feature Extraction}
Each $S^m$ is processed by a modality-specific semantic encoder, parameterized by $\delta^m$, to yield a feature vector $X^m = \delta^m(S^m)$. For this purpose, we use pre-trained deep neural networks (NNs), such as BERT for text and transformer-based architectures for image and audio, to extract high-quality semantic features.

\subsubsection{Uni-modal VIB (U-VIB)}
To compress each uni-modal representation and preserve only task-relevant information, $X^m$ is passed through a U-VIB module, parameterized by $\varepsilon^m$. This module outputs a representation $Z^m = \varepsilon^m(X^m)$, by reducing intra-modal redundancy. The U-VIB is implemented by a uni-modal variational autoencoder (VAE). 

\subsubsection{Multi-modal Fusion and MI reduction}
The set of uni-modal latent variables $\{Z^i, Z^t, Z^a\}$ is first aggregated by a fusion function (e.g., concatenation) to form a joint representation. Subsequently, a cross-modal MI redundancy suppression network (i.e., $\zeta^{i,t}, \zeta^{i,a}, \zeta^{t,a}$ ) processes the representation to reduce inter-modal redundancies, producing the efficient latent representation $X$ for further multi-modal processing. More details will be given in Section \ref{sec:MI}.

\subsubsection{Multi-modal Encoder}
The multi-modal representation $X$ further goes through a multi-modal encoder NN parameterized by $\eta$ to produce a robust feature $Z$ for channel transmission. This module is jointly trained as part of the latter multi-modal VIB.

\subsection{Channel Model}
The multi-modal representation $Z$ is transmitted over wireless channels. The received signal, $\hat{Z}$, is
\begin{equation}
    \hat{Z} = h(Z) + \epsilon
    \label{eq:channel},
\end{equation}
where $h(\cdot)$ represents the channel transfer function (e.g., channel gains with fading), and $\epsilon$ is the additive white Gaussian noise (AWGN).


\subsection{Receiver Architecture}
The goal of the receiver is to reconstruct the task-relevant information against channel distortion.

\subsubsection{Multi-modal VIB (M-VIB)}
At the receiver side, an M-VIB is applied to refine the perturbed signal $\hat{Z}$, and recover the transmitted semantic information to generate the output $\hat{Y}$ using NN  $\upsilon$. This module is jointly trained with the previous multi-modal encoder NN $\eta$ as the M-VIB network. The M-VIB module is implemented by a multi-modal VAE.

The whole process is denoted as follows: 
\begin{equation}
    S^m \xrightarrow{\delta^m} X^m \xrightarrow{\varepsilon^m} Z^m \xrightarrow{\zeta} X \xrightarrow{\eta} Z \xrightarrow{channel} \hat{Z} \xrightarrow{\upsilon} \hat{Y}.
\end{equation}

\section{Two-stage Variational Information Bottleneck}
\label{sec:VIB}

This section provides a detailed introduction to our proposed two-stage multi-modal VIB framework, based on the VIB principle. In Stage I (Section~\ref{sec:uvib}), each modality is processed independently via a U-VIB to obtain compact, task-relevant representations. In Stage II (Section~\ref{sec:mvib}), the compressed uni-modal representations are first fused and then passed through M-VIB, which yields an efficient, robust representation against channel impairments.

\subsection{Uni-modal Variational Information Bottleneck}
\label{sec:uvib}

\begin{figure}[t]
\centering
\includegraphics[width=0.8\linewidth]{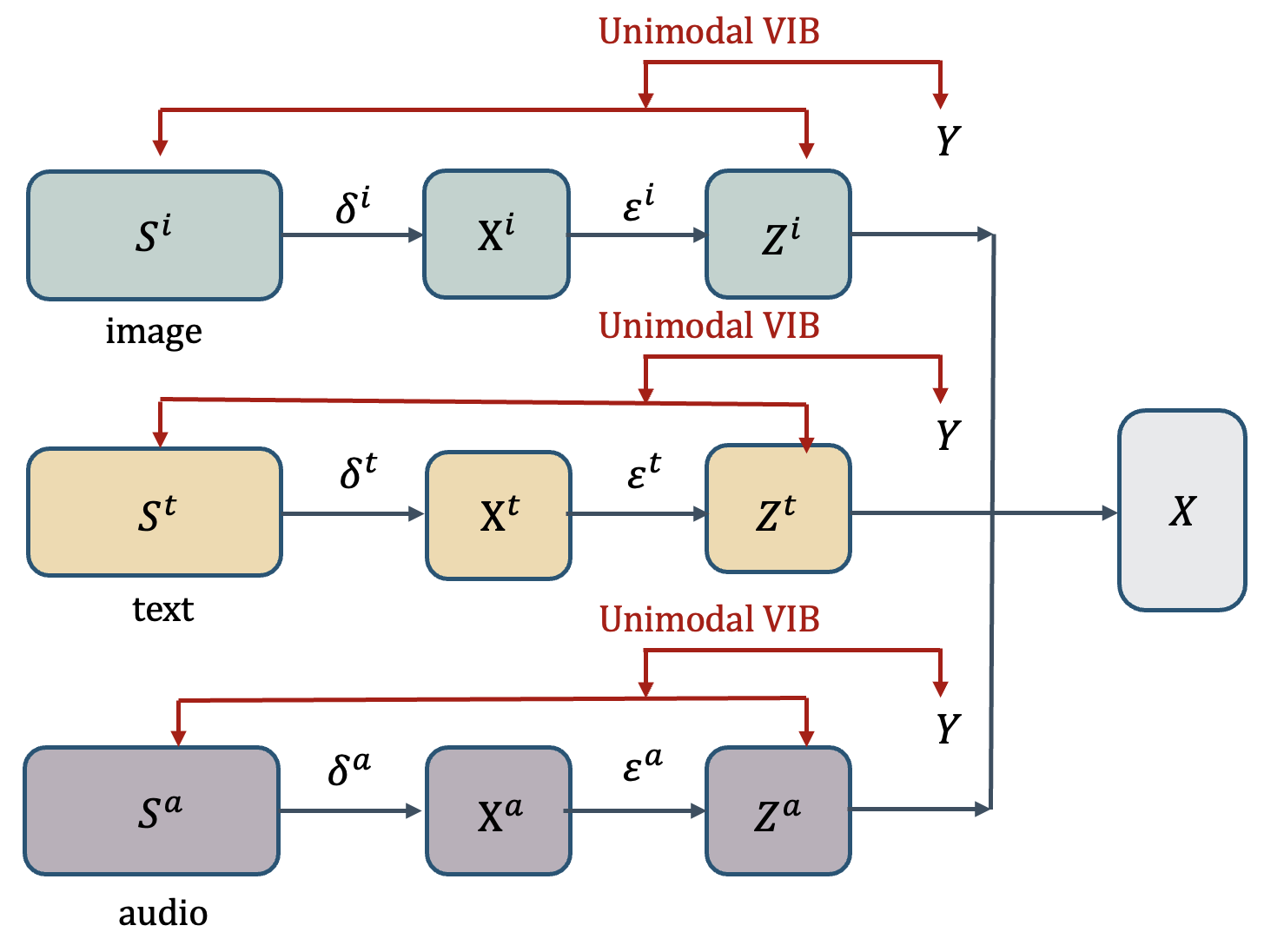}
\caption{Uni-modal VIB.}
\label{uni}
\end{figure}

We will first give details on the compression process of each modality into a stochastic latent code $Z^m$, $m \in \{i, t, a\}$, using the proposed U-VIB. The objective is to create the minimal representation by removing redundant information from the raw data $S^m$ while preserving essential information about the target $Y$. We first derive the objective and then describe its implementation using lightweight, modality-specific encoders. The architecture of the proposed U-VIB is illustrated in Fig.~\ref{uni}. 

For a single modality \(m \in \{ i, t, a \}\), the objective is to optimize the trade-off between data compression and task performance, i.e., 
\begin{equation}
\label{tradeoff_obj_uni}
    \min_{p(z^m \vert s^m)}\; I(S^m; Z^m) - \beta\, I(Z^m; Y),
\end{equation}
where $\beta\!>\!0$ balances data compression $ I(S^m; Z^m)$ against task performance $I(Z^m ; Y)$. Solving \eqref{tradeoff_obj_uni} directly is computationally prohibitive. We omit the intermediate variable $X^m$ to simplify the illustration. Following the derivation in Appendix A, we obtain an upper bound for $ I(S^m; Z^m)$,
\begin{equation}
    I(S^m; Z^m) \leq \int p(s^m)\,p(z^m \vert s^m) \log \frac{p(z^m \vert s^m)}{q(z^m)}\,ds^m\,dz^m,
    \label{eq:compression}
\end{equation}
where $q(z^m)$ is variational approximation to the true marginal distribution $p(z^m)$.
Similarly, the predictive MI $I(Z^m; Y)$ is lower-bounded using an auxiliary decoder $q(y \vert z^m)$,
\begin{align}
    I(Z^m; Y) \geq\; & \int p(s^m)\,p(y \vert s^m)\,p(z^m \vert s^m)\, \nonumber \\
    & \log q(y \vert z^m)\, ds^m\,dy\,dz^m.
    \label{eq:prediction}
\end{align}

Combining \eqref{eq:compression} and \eqref{eq:prediction}, a lower bound of \eqref{tradeoff_obj_uni} is
\begin{align}
    & I(S^m; Z^m) - \beta\, I(Z^m; Y) \nonumber \\
    &\geq 
    \mathbb{E}_{p(s^m)} \!\left[
        D_{\mathrm{KL}}\!\big( p(z^m|s^m) \,\|\, q(z^m) \big)
    \right] \notag \\
    &\quad
    - \beta\, \mathbb{E}_{p(s^m, y)\, p(z^m|s^m)} \!\left[
        \log q(y|z^m)
    \right].
\end{align}
 
Using uni-modal encoder networks parameterized by $\delta^m$ and $\varepsilon^m$ to solve the problem in \eqref{tradeoff_obj_uni}, the objective function is reformulated as 
\begin{align}   \mathcal{L}_{\text{U-VIB}} =\; & \mathbb{E}_{p(s^m)}\left[ D_{\mathrm{KL}}\big(p(z^m \vert s^m) \,\|\, q(z^m)\big) \right] \nonumber \\
    & - \beta\, \mathbb{E}_{p(s^m, y)}\left[ \mathbb{E}_{p(z^m \vert s^m)}[\log q(y \vert z^m)] \right].
    \label{eq:vib_loss}
\end{align}
\begin{figure}[t]
\centering
\includegraphics[width=0.8\linewidth]{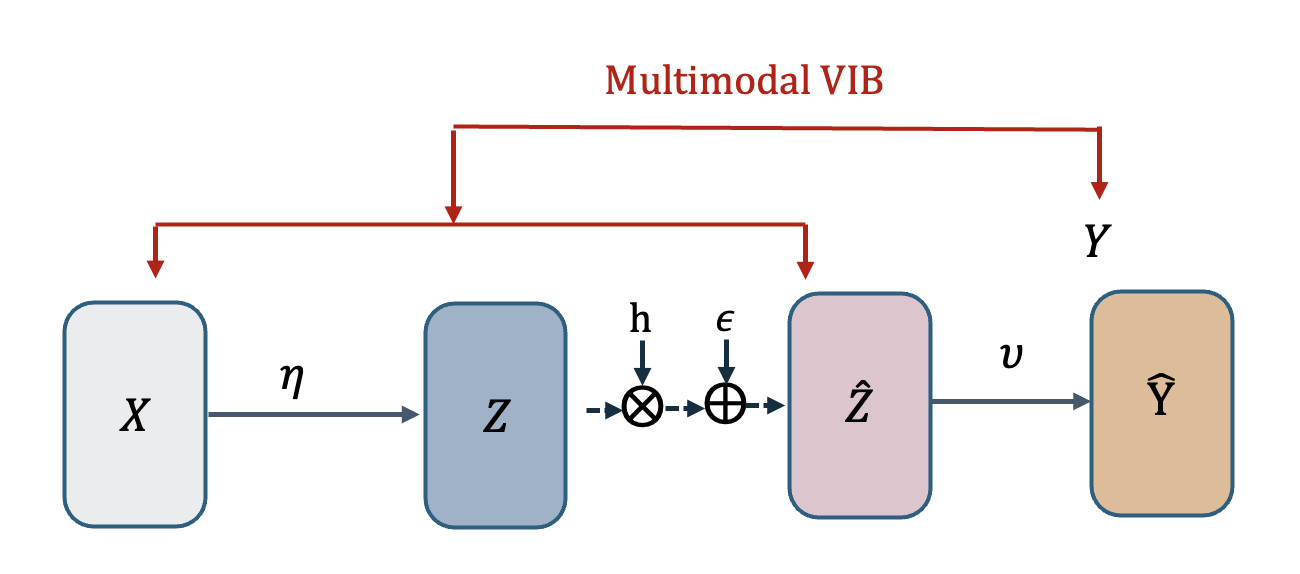}
\caption{Multi-modal VIB.}
\label{multi}
\end{figure}
The conditional latent distribution $p(z^m \vert s^m)$ normally follows a multi-variate Gaussian \cite{alemi2016deep, wang2019deep}, of which mean and covariance are learned via NNs,
\begin{equation}
    p(z^m \vert s^m) = \mathcal{N}(\mu_{z^m}, \Sigma_{z^m}).
    \label{eq:gaussian_encoder_1}
\end{equation}

To enable gradient-based optimization, we adopt the reparameterization method \cite{alemi2016deep, wang2019deep}:
\begin{equation}
    z^m = \mu_{z^m} + \Sigma_{z^m} \cdot \epsilon, \quad \epsilon \sim \mathcal{N}(0, I),
    \label{eq:reparam}
\end{equation}
which transfers the stochasticity to an auxiliary noise variable $\epsilon$, allowing gradients to propagate through $\mu_z$ and $\Sigma_z$. For classification tasks, the predictive likelihood $q(y \vert z^m)$ is calculated using cross-entropy loss, i.e., 
\begin{equation}
    \log q(y \vert z^m) = y \log \hat{y}, \quad \hat{y} = \mathrm{Sigmoid}(Dec(z^m)),
    \label{eq:classification}
\end{equation} 
where $Dec(z^m)$ denotes the decoder used for each $Z^m$, and a Sigmoid function is applied for classification. The prior distribution $q(z^m)$ is fixed to be Gaussian~\cite{alemi2016deep, wang2019deep} in practice, i.e., 
\begin{equation}
    q(z^m) = \mathcal{N}(0, I).
    \label{eq:prior}
\end{equation}
Given \eqref{eq:gaussian_encoder_1} and \eqref{eq:prior}, the KL divergence is computed in a closed form as
\begin{equation}
    D_{\mathrm{KL}}\big(p(z^m \vert s^m) \,\|\, q(z^m)\big)
    = D_{\mathrm{KL}}\big(\mathcal{N}(\mu_{z^m}, \Sigma_{z^m}) \,\|\, \mathcal{N}(0, I)\big).
    \label{eq:kl}
\end{equation}
Combining \eqref{eq:classification} and \eqref{eq:kl},  $L_{\text{U-VIB}}$ can be approximated via Monte Carlo sampling over a mini-batch of samples with size $n$,
\begin{align}
    \mathcal{L}_{\text{U-VIB}} \approx\; & \frac{1}{n} \sum_{i=1}^{n}
    \log q(y_i \vert z^m_i) \nonumber \\
    & - \beta \sum_{i=1}^{n}
    D_{\mathrm{KL}}\left( \mathcal{N}(\mu_{z^m_i}, \Sigma_{z^m_i}) \,\|\, \mathcal{N}(0, I) \right).
    \label{eq:loss_uvib}
\end{align}

 The Uni-modal loss $\mathcal{L}_{\text{U-VIB}}$ encourages representations $Z^m$ to retain the maximal information about target $Y$ while suppressing irrelevant details from input $S^m$ through KL-based regularization. In addition to being computationally feasible,  $\mathcal{L}_{\text{U-VIB}}$ formulation has two primary advantages. First, the KL-based regularization in \eqref{eq:loss_uvib} enforces principled compression by pulling $p(z^m \mid s^m)$ towards a simple prior \(q(z^m)\) (e.g., \(\mathcal{N}(0,I)\)). This constraint reduces modality-specific noise in \(Z^m\). Second, it ensures task sufficiency by preserving information only related to the target $Y$. Consequently, the U-VIB module produces compact and task-relevant uni-modal embeddings, which serve as effective inputs for the subsequent multi-modal fusion process.

\subsection{Multi-modal Variational Information Bottleneck}
\label{sec:mvib}

In the following, we explain how the multi-modal representation $X$ is compressed into the minimal but sufficient representation $\hat{Z}$ with robustness against channel distortion. Similar to U-VIB, our objective is to optimize the trade-off between data compression and task performance,

\begin{equation}
\label{tradeoff_obj_multi}
    \min_{p(\hat{z} \vert x)}\; I(X; \hat{Z}) - \gamma\, I(\hat{Z}; Y),
\end{equation}
where $\gamma\!>\!0$ balances compression $ I(X; \hat{Z})$ against task relevance $I(\hat{Z}; Y)$. Similar to U-VIB, for computationally feasible, $ I(X; \hat{Z})$ can be upper-bounded by
\begin{equation}
    I(X; \hat{Z}) \leq \int p(x)\,p(\hat{z} \vert x) \log \frac{p(\hat{z} \vert x)}{q(\hat{z})}\,dx\,d\hat{z},
    \label{eq:compression2}
\end{equation}
where $q(\hat{z})$ is a variational approximation to the marginal distribution $p(\hat{z})$. 
The predictive MI $I(\hat{Z}; Y)$ is lower-bounded by an auxiliary decoder $q(y \vert \hat{z})$,
\begin{align}
    I(\hat{Z}; Y) \geq\; & \int p(\hat{z})\,p(y \vert \hat{z})\,p(\hat{z} \vert x)\, \nonumber \\
    & \log q(y \vert \hat{z})\, dx\,dy\,d\hat{z}.
    \label{eq:prediction2}
\end{align}
Combining \eqref{eq:compression2} and \eqref{eq:prediction2}, we can lower bound  \eqref{tradeoff_obj_multi} by
\begin{align}
    I(X; \hat{Z}) - \gamma\, I(\hat{Z}; Y)
    &\geq 
    \mathbb{E}_{p(x)}\left[ D_{\mathrm{KL}}\big(p(\hat{z} \vert x) \,\|\, q(\hat{z})\big) \right] \nonumber \\
    & - \gamma\, \mathbb{E}_{p(x, y)}\left[ \mathbb{E}_{p(\hat{z} \vert x)}[\log q(y \vert \hat{z})] \right].
\end{align}

Using a multi-modal encoder network parameterized by $\eta$ to solve  optimization problem (\ref{tradeoff_obj_multi}), the objective becomes 
\begin{align}
    \mathcal{L}_{\text{M-VIB}} =\; & \mathbb{E}_{p(x)}\left[ D_{\mathrm{KL}}\big(p(\hat{z} \vert x) \,\|\, q(\hat{z})\big) \right] \nonumber \\
    & - \gamma\, \mathbb{E}_{p(x, y)}\left[ \mathbb{E}_{p(\hat{z} \vert x)}[\log q(y \vert \hat{z})] \right].
    \label{eq:vib_loss2}
\end{align}

 Similar to U-VIB, to minimize the loss $L_{\text{M-VIB}}$ using deep NNs, we assume that the conditional latent distribution $p(\hat{z} \vert x)$ follows a multi-variate Gaussian distribution \cite{alemi2016deep, wang2019deep}, whose mean and covariance are learned via NNs,
\begin{equation}
    p(\hat{z} \vert x) = \mathcal{N}(\mu_{\hat{z}}', \Sigma_{\hat{z}}').
    \label{eq:gaussian_encoder}
\end{equation}
The reparameterization method \cite{alemi2016deep, wang2019deep} is used to enable gradient-based optimization of the model parameters. For classification tasks, the predictive likelihood $q(y \vert z)$ is calculated using cross-entropy loss, i.e., 
\begin{equation}
    \log q(y \vert \hat{z}) = y \log \hat{y}, \quad \hat{y} = \mathrm{Sigmoid}(Dec(\hat{z})),
    \label{eq:classification2}
\end{equation} 
where $Dec(\hat{z})$ denotes the decoder used for each $\hat{Z}$, and a Sigmoid function is applied for classification. The prior distribution $q(\hat{z})$ is fixed to be Gaussian~\cite{alemi2016deep, wang2019deep}, i.e., 
\begin{equation}
    q(\hat{z}) = \mathcal{N}(0, I).
    \label{eq:prior2}
\end{equation}

Then, KL divergence is obtained in a closed form as
\begin{equation}
D_{\mathrm{KL}}\big(p(\hat{z} \vert x) \,\|\, q(\hat{z})\big)
    = D_{\mathrm{KL}}\big(\mathcal{N}(\mu_{\hat{z}}', \Sigma_{\hat{z}}') \,\|\, \mathcal{N}(0, I)\big).
    \label{eq:kl2}
\end{equation}
Combining \eqref{eq:classification2} and \eqref{eq:kl2}, the multi-modal loss function $\mathcal{L}_{\text{M-VIB}}$ can be approximated using Monte Carlo sampling over a mini-batch of samples with size $n$,
\begin{align}
    \mathcal{L}_{\text{M-VIB}} \approx\; & \frac{1}{n} \sum_{i=1}^{n}
    \log q(y_i \vert \hat{z_i}) \nonumber \\
    & - \gamma \sum_{i=1}^{n}
    D_{\mathrm{KL}}\left( \mathcal{N}(\mu_{\hat{z_i}}', \Sigma_{\hat{z_i}}') \,\|\, \mathcal{N}(0, I) \right).
    \label{eq:loss_mvib}
\end{align}

The multi-modal loss $\mathcal{L}_{\text{M-VIB}}$ encourages the multi-modal representation $\hat{Z}$ to preserve the needed information about the target $Y$ while compressing irrelevant information from the input multi-modal representation $X$ via KL-based regularization. The trade-off between task relevance and compression is controlled by the coefficient $\gamma$. Using $\mathcal{L}_{\text{M-VIB}}$ formulation has two major benefits. First, compression is achieved by using the KL regularization in \eqref{eq:loss_mvib}, which constrains \(p(\hat{z}\mid x)\) to remain close to a simple prior \(q(\hat{z})\) (e.g., \(\mathcal{N}(0, I)\)), and simultaneously producing latent representations that are more robust to channel noise. Second, it ensures task sufficiency by preserving only the predictive information required for the target $Y$. As a result, the multi-modal VIB produces compact and task-relevant representations that enable reliable transmission in multi-modal scenarios.

\section{Cross-modal Redundancy Reduction via MI Minimization}
\label{sec:MI}

The fused multi-modal representation $X$ is informative, but still contains redundancy across modalities. Therefore, we propose the cross-modal redundancy reduction using MI minimization. In what follows, we  first derive the MI minimization module among different modalities. Then we show the NN realization of this module.

\subsection{Redundancy Reduction via MI Minimization}

\begin{figure}[t]
\centering
\includegraphics[width=0.7\linewidth]{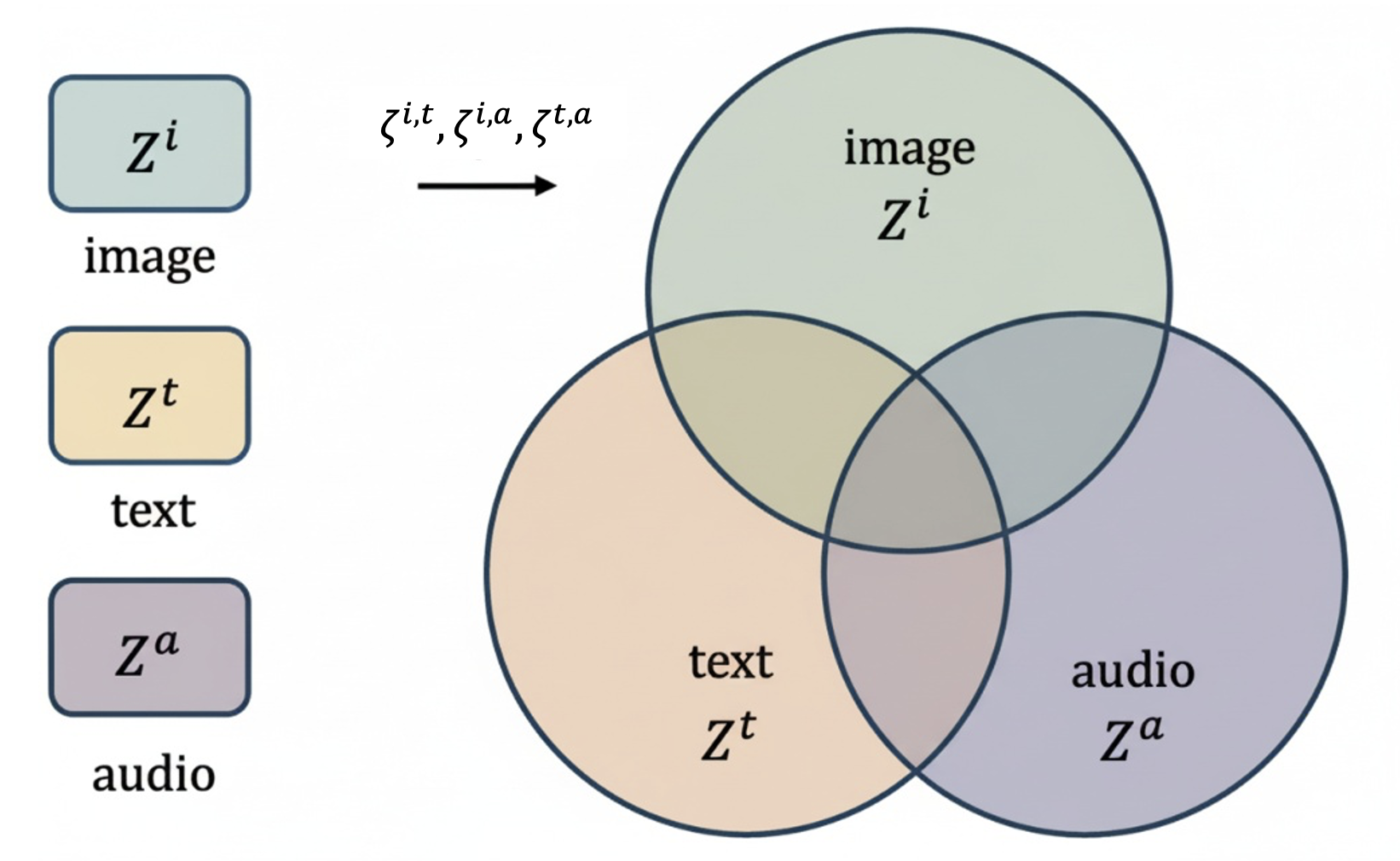}
\caption{Redundancy among different modalities.}
\label{mi}
\end{figure}

To reduce the redundancy across three modalities $Z^i, Z^t, Z^a$, we should minimize their joint MI, 
\begin{equation}
I(Z^i, Z^t, Z^a) \;=\;
D_{\text{KL}}\!\bigl(p_{z^i z^t z^a} \,\big\|\, p_{z^i}p_{z^t}p_{z^a}\bigr).
\end{equation}
However, directly minimizing $I(z^i, z^t, z^a)$ is impractical, as it is non-trivial to obtain a trackable $I(Z^i, Z^t, Z^a)$ for high-dimensional variables.  
Consequently, we consider the redundancy of the \emph{sum of pair-wise} MIs, which admit reliable variational bounds and are efficient for suppressing redundancy across modalities.
To minimize the MI between each modality pair, the sum redundancy loss is defined as,
\begin{equation}
\mathcal{L}_{\text{red}} = I(Z^i;Z^t) + I(Z^i;Z^a) + I(Z^t;Z^a),
\label{eq:redundancy_loss}
\end{equation}

\noindent where the MIs between random variables are given by the KL divergence as
\begin{align}
I(Z^i;Z^t) &= D_{\text{KL}}(p_{z^i z^t} \,\|\, p_{z^i} p_{z^t}), \\
I(Z^i;Z^a) &= D_{\text{KL}}(p_{z^i z^a} \,\|\, p_{z^i} p_{z^a}), \\
I(Z^t;Z^a) &= D_{\text{KL}}(p_{z^t z^a} \,\|\, p_{z^t} p_{z^a}).
\end{align}

While pair-wise MI simplifies the problem, estimating KL divergence of modality pairs directly in high-dimensional spaces is still challenging. To address this problem, we establish a tractable variational lower bound on the MI. We begin by relating MI to the JS divergence. The JS divergence between two distributions, in our case $p_{z^i z^t}$ and $p_{z^i} p_{z^t}$, is defined as:
\begin{equation}
D_{\text{JS}}(p_{z^i z^t} \,\|\, p_{z^i} p_{z^t}) = \frac{1}{2} D_{\text{KL}}(p_{z^i z^t} \,\|\, m) + \frac{1}{2} D_{\text{KL}}(p_{z^i} p_{z^t} \,\|\, m),
\end{equation}
where $m = \frac{1}{2}(p_{z^i z^t} + p_{z^i} p_{z^t})$. Then we have

\begin{equation}
\begin{aligned}
2D_{\text{JS}}(p_{z^i z^t} \,\|\, p_{z^i} p_{z^t})
&= D_{\text{KL}}(p_{z^i z^t} \,\|\, m)
 + D_{\text{KL}}(p_{z^i} p_{z^t} \,\|\, m) \\
&= \underbrace{D_{\mathrm{KL}}(p_{z^i z^t}\,\|\,p_{z^i}p_{z^t})}_{=\,I(Z^i;Z^t)} \\
&\quad + \mathbb{E}_{p_{z^i z^t}}
   \!\Big[\log \tfrac{p_{z^i} p_{z^t}}{m}\Big] \\
&\qquad + \mathbb{E}_{p_{z^i} p_{z^t}}
   \!\Big[\log \tfrac{p_{z^i} p_{z^t}}{m}\Big] \\
&= I(Z^i;Z^t)
   + 2\,\mathbb{E}_{m}\!\Big[\log \tfrac{p_{z^i} p_{z^t}}{m}\Big] \\
&= I(Z^i;Z^t)
   - 2\,D_{\mathrm{KL}}(m\,\|\,p_{z^i} p_{z^t}).
\end{aligned}
\label{eq:js_decomp}
\end{equation}

Thus, we can lower bound the MI as
\begin{equation}
\begin{aligned}
\label{eq:mi_js_bound}
I(Z^i;Z^t)
&= 2D_{\mathrm{JS}}(p_{z^i z^t} \,\|\, p_{z^i} p_{z^t})
   + 2D_{\mathrm{KL}}(m\|p_{z^i} p_{z^t}) \\
&\;\ge\; 2D_{\mathrm{JS}}(p_{z^i z^t} \,\|\, p_{z^i} p_{z^t}) \,.
\end{aligned}
\end{equation}
The equality holds if and only if (\emph{iff})
\(D_{\mathrm{KL}}(m\,\|\,p_{z^i} p_{z^t})=0\), i.e., \(m=p_{z^i}p_{z^t}\), which implies $p_{z^i z^t}=p_{z^i}p_{z^t}$, and $Z^i$ and $Z^t$ are independent. 
The bound is advantageous because JS divergence  variational representation derived from the f-GAN framework \cite{nowozin2016f, hjelm2019learning}, with which the JS divergence can be estimated by training a discriminator network $T(\cdot)$, to distinguish samples from the joint distribution (positive pairs) from samples from the product of marginal distributions (negative pairs), i.e.,
\begin{align} \label{JSDivergence} 
D_{\text{JS}}(p_{z^i z^t} \,\|\, p_{z^i} p_{z^t})
&= \frac{1}{2}\sup_{T_{\text{it}} \in \mathcal{F}} \Big[\,
 \mathbb{E}_{ p_{z^i,z^t}} \big[ \log \sigma(T_{\text{it}}(z^i,z^t)) \big] \nonumber \\
& +\, 
 \mathbb{E}_{p_{z^i} p_{z^t}} \big[ \log (1 - \sigma(T_{\text{it}}(z^i,z^t)))\Big] \Big] \nonumber  \\
& +\log 2,
\end{align}
where \( T_{\text{it}} \) is a discriminator between modality $i$ and modality $t$ and trained to distinguish between samples drawn from \(p_{z^i z^t}  \). \( p_{z^i} p_{z^t} \).
\( \sigma(\cdot) \) is the sigmoid function. With (\ref{JSDivergence}) as the objective function, \( \mathbb{E}_{ p_{z^i,z^t}} \big[ \log \sigma(T_{\text{it}}(z^i,z^t))\big] \) encourages the discriminator $T_{\text{it}}$ to assign high confidence to samples from \( p_{z^i,z^t} \), while \( \mathbb{E}_{ p_{z^i} p_{z^t}} \big[ \log (1 - \sigma(T_{\text{it}}(z^i,z^t))) \big] \) encourages the discriminator $T_{\text{it}}$ to assign low confidence to samples from \( p_{z^i} p_{z^t} \). The supremum over \( T_{\text{it}} \in \mathcal{F} \) seeks the optimal discriminator that maximizes $D_{\text{JS}}(p_{z^i z^t} \,\|\, p_{z^i} p_{z^t})$. The constant \( \log 2 \) ensures consistency with the standard definition of JS divergence. By combining the bound in \eqref{eq:mi_js_bound} with the variational form in \eqref{JSDivergence}, we obtain a variational lower bound for the MI:
\begin{equation}
\begin{aligned}
I(Z^i;Z^t)
&\ge \sup_{T_{\text{it}} \in \mathcal{F}} \Big\{
    \mathbb{E}_{p_{z^i z^t}}
      \log \sigma\!\big(T_{\mathrm{it}}(z^i,z^t)\big) \\
&\quad + \mathbb{E}_{p_{z^i}p_{z^t}}
      \log\!\Big(1-\sigma\!\big(T_{\mathrm{it}}(z^i,z^t)\big)\Big)
\Big\} \\
&\quad + 2\log 2 .
\end{aligned}
\label{eq:mi_lower_bound}
\end{equation}
The supremum $\sup_{T_{\text{it}} \in \mathcal{F}}$ indicates that the bound holds for the optimal discriminator chosen from a function class $\mathcal{F}$. While the theoretical bound is tight, finding the true supremum across an entire function class is intractable. In practice, we approximate the supremum by parameterizing the discriminator as an NN $T_{\mathrm{it}}(z^i,z^t;\zeta^{i,t})$, where $\zeta^{i,t}$ is the trainable parameter in the network. Approximating with NNs transforms the problem into a practical, gradient-based optimization problem. We define the objective $\mathcal{J}_{\log\sigma}(Z^i;Z^t)$, which denotes the bound for a specific, parameterized discriminator: 
\begin{align}\label{eq:logsigmoid_mi}
\mathcal{J}_{\log\sigma}(Z^i;Z^t)
&= \mathbb{E}_{p_{z^i,z^t}}
   \Big[\log \sigma\!\big(T_{\text{it}}(z^i,z^t;\zeta^{i,t})\big)\Big] \nonumber\\
&\quad + \mathbb{E}_{p_{z^i} p_{z^t}}
   \Big[\log \!\big(1 - \sigma\!\big(T_{\text{it}}(z^i,z^t;\zeta^{i,t})\big)\big)\Big]
   \nonumber \\
&\quad + 2\log 2 .
\end{align}
The connection between the practical objective and the theoretical supremum is realized through adversarial training. The discriminator network \( T_{\text{it}} \) is trained to maximize $\mathcal{J}_{\log\sigma}$, thereby driving it to approximate the optimal discriminator. Concurrently, the encoders that generate the representations $z^i$ and $z^t$ are trained to minimize $\mathcal{J}_{\log\sigma}$, which will be detailed in Section \ref{sec:grl}. The following proposition characterizes the behavior of $\mathcal{J}_{\log\sigma}$.

\begin{proposition} \label{Prop_1}
For the optimal discriminator $T_{\mathrm{it}}$, the objective (loss function) in \eqref{eq:logsigmoid_mi} satisfies
\[
0 \;\le\; \mathcal{J}_{\log\sigma}(Z^i;Z^t) \;\le\; 2\log 2,
\]
with $\mathcal{J}_{\log\sigma}(Z^i;Z^t)=0$ iff $p_{z^i,z^t}=p_{z^i}p_{z^t}$, while the upper bound $2\log 2$ is attained in the limit of perfectly separable distributions $p_{z^i,z^t}$ and $p_{z^i}p_{z^t}$.
\end{proposition}
The proof is provided in Appendix \ref{proof_2log2}.

\begin{algorithm}[t]
\caption{Training process of the proposed multi-modal TOC framework}
\label{alg:cmib-train}
\textbf{Input} Training set $\mathcal{S}_{\text{train}}$, epochs $T$, batch size $B$, channel coefficient $h$, noise level $\sigma^2$, and redundancy coefficient $\lambda_{\text{red}}$ \\
\textbf{Output} Trained network with parameters $\delta^m, \varepsilon^m, \zeta^{i,t}, \zeta^{i,a}, \zeta^{t,a}, \eta, \upsilon$ \\
\textbf{Initialization} $t \gets 1$, GRL scale $\alpha \gets 0$
\begin{algorithmic}[1]
\While{$t \leq T$}
    \State Sample a mini-batch $\{(s^i_j,s^t_j,s^a_j,y_j)\}_{j=1}^{B} \subset \mathcal{S}_{\text{train}}$
           \State $X^m \gets \delta^m(S^m)$, \quad ($\forall m\in\mathcal{M}$)  
            \State $Z^m\gets \varepsilon^m(X_m)$ 
           \State $(\mu_m,\sigma_m)\gets \text{Encoder}_m(Z_m)$
           \State $\hat{y}_m\gets \text{Decoder}_m(z_m)$
           \State $\text{Compute }  \mathcal{L}_{\text{U-VIB}} $ 
           \State $X \gets \zeta(Z_i,Z_t,Z_a)$
           \State $\text{Compute }  \mathcal{L}_{\text{red}}$
           \State $Z \gets \eta(X)$
       \State $\hat{Z} \gets h(Z) + \epsilon,\quad \epsilon\sim\mathcal{N}(0,\sigma^2 I)$
           \State $(\mu,\sigma)\gets \text{Encoder}(\hat{Z})$
           \State \ $\hat{y}\gets \text{Decoder}(\hat{z})$
           \State $\text{Compute }  \mathcal{L}_{\text{M-VIB}}$
           \State Compute $\mathcal{L} \gets \sum_{m \in \{i,t,a\}}\mathcal{L}_{\text{U-VIB}} + \mathcal{L}_{\text{M-VIB}} + \lambda_{\text{red}}\mathcal{L}_{\text{red}}$ 
    \State $\text{Update all trainable modules accordingly} $
   
\EndWhile
\end{algorithmic}
\end{algorithm}

Proposition \ref{Prop_1} shows that our objective $\mathcal{J}_{\log\sigma}$ provides a bounded and meaningful measure of redundancy. When we train the NN to minimize $\mathcal{J}_{\log\sigma}$, we are explicitly pushing the representations towards independence (i.e., $\mathcal{J}_{\log\sigma} = 0$). The bounded nature of this objective is also highly desirable, as it enhances the stability of the adversarial training process by preventing exploding gradients.

Similarly, we define objectives for the other modality pairs. The objective between modality $i$ and modality $a$ is
\begin{align}\label{eq:logsigmoid_mi_2}
\mathcal{J}_{\log\sigma}(Z^i;Z^a)
&= \mathbb{E}_{p_{z^i,z^a}}
   \big[\log \sigma\!\big(T_{\text{it}}(z^i,z^a;\zeta^{i,a})\big)\big] \nonumber\\
&\quad + \mathbb{E}_{p_{z^i} p_{z^a}}
   \big[\log \big(1 - \sigma\!\big(T_{\text{it}}(z^i,z^a;\zeta^{i,a})\big)\big)\big] \nonumber\\
&\quad + 2\log 2 ,
\end{align}
where $T_{\text{ia}}(\cdot)$ is the discriminator between modality $i$ and  $a$. The objective between modality $t$ and  $a$ is
\begin{align}\label{eq:logsigmoid_mi_3}
\mathcal{J}_{\log\sigma}(Z^t;Z^a)
&= \mathbb{E}_{p_{z^t,z^a}}
   \big[\log \sigma\!\big(T_{\text{ta}}(z^t,z^a;\zeta^{t,a})\big)\big] \nonumber\\
&\quad + \mathbb{E}_{p_{z^t} p_{z^a}}
   \big[\log \big(1 - \sigma\!\big(T_{\text{ta}}(z^t,z^a;\zeta^{t,a})\big)\big)\big] \nonumber \\
&\quad + 2\log 2 ,
\end{align}
where $T_{\text{ta}}(\cdot)$ is the discriminator between modality $t$ and $a$. By substituting these tractable, bounded, and differentiable objectives in \eqref{eq:logsigmoid_mi}, \eqref{eq:logsigmoid_mi_2}, and \eqref{eq:logsigmoid_mi_3}  to our original formulation in \eqref{eq:redundancy_loss}, the final redundancy becomes the sum of these pairwise objectives (loss functions),
\begin{equation}
\mathcal{L}_{\text{red}} = \mathcal{J}_{\log\sigma}(Z^i;Z^t) + \mathcal{J}_{\log\sigma}(Z^i;Z^a) + \mathcal{J}_{\log\sigma}(Z^t;Z^a).
\label{eq:loss_red}
\end{equation}
Minimizing $\mathcal{L}_{\text{red}}$ via end-to-end training effectively encourages the model to learn complementary representations by penalizing the statistical dependency between each pair of modalities. We note that each $\mathcal{J}_{\log\sigma}$ is a lower bound on MI, which is tightened by training a corresponding discriminator to maximize its value. Therefore, the optimization process is inherently adversarial. In the following section, we detail the adversarial training strategy to minimize $\mathcal{L}_{\text{red}}$.

\subsection{Gradient Reversal Layer for Adversarial Redundancy Suppression}
\label{sec:grl}

As discussed above, to minimize loss $\mathcal{L}_{\text{red}}$, we will minimize the MI of modality pairs by an adversarial learning approach, in which the encoders are trained to produce representations indistinguishable to a set of discriminators. The process leads to a min-max objective over the redundancy loss:
\begin{equation}
\min_{\text{encoder}} \max_{T_{\text{it}},\, T_{\text{ia}},\, T_{\text{ta}}} 
\mathcal{L}_{\text{red}}.
\label{eq:minmax}
\end{equation}
 The \emph{min} operation in \eqref{eq:minmax} is to force the representations for an encoder to become less redundant. Simultaneously, the discriminators $T_{\text{it}}$, $T_{\text{ia}}$, and $T_{\text{ta}}$ (parameterized by $\zeta^{\text{it}}$, $\zeta^{\text{ia}}$ and $\zeta^{\text{ta}}$) are trained to maximize $\mathcal{L}_{\text{red}}$ by making each $\mathcal{J}_{\log\sigma}$ a tighter and higher estimate of the true MI.

A common solution for such min-max problems is through alternative optimization (AO) of the encoder and the discriminator, where one set of parameters is updated while the others held fixed. However, AO could be unstable and computationally inefficient. To achieve a more efficient and stable single-pass training strategy, we adopt the GRL \cite{ganin2016domain}, which is a special layer that acts as an identity function during the forward pass but reverses and scales the gradient during the backward pass. For a given feature vector $\mathbf{h}$ and a hyper-parameter $\alpha > 0$, GRL is defined as:
\begin{equation}
\mathrm{GRL}_\alpha(\mathbf{h}) =
\begin{cases}
\mathbf{h}, & \text{(forward pass)},\\[2pt]
-\alpha\,\partial\mathcal{L}/\partial\mathbf{h}, & \text{(backward pass)}.
\end{cases}
\label{eq:grl_alpha}
\end{equation}
The GRL layer is placed between an encoder and a discriminator to implement the min-max operation. During backpropagation, the discriminator receives the standard gradient to maximize its objective (e.g., $\mathcal{J}_{\log\sigma}$). However, the gradient that flows back to the encoder is reversed, causing the encoder to update its parameters in the opposite direction, thus minimizing the objective of the discriminators. In our NN architecture, we insert a GRL before passing the encoded representations to their respective discriminators. Accordingly, the modality representations $Z_i$, $Z_t$, and $Z_a$ go through their respective discriminators through GRLs as
\begin{align}
T_{\text{it}} &= T_{\text{it}}\bigl(\mathrm{GRL}_\alpha(Z_i), Z_t\bigr), \\
T_{\text{ia}} &= T_{\text{ia}}\bigl(\mathrm{GRL}_\alpha(Z_i), Z_a\bigr), \\
T_{\text{ta}} &= T_{\text{ta}}\bigl(\mathrm{GRL}_\alpha(Z_t), Z_a\bigr).
\end{align}
With GRL, $\mathcal{L}_{\text{red}}$ in \eqref{eq:minmax} can be minimized with respect to the encoder and maximized with respect to $T_{\text{it}}$, $T_{\text{ia}}$, and $T_{\text{ta}}$ within a single backward process.

\begin{figure*}[t]
  \centering
  \newcommand{\imgw}{0.23\linewidth}

  \begin{subfigure}[t]{\imgw}
    \centering
    \includegraphics[width=\linewidth]{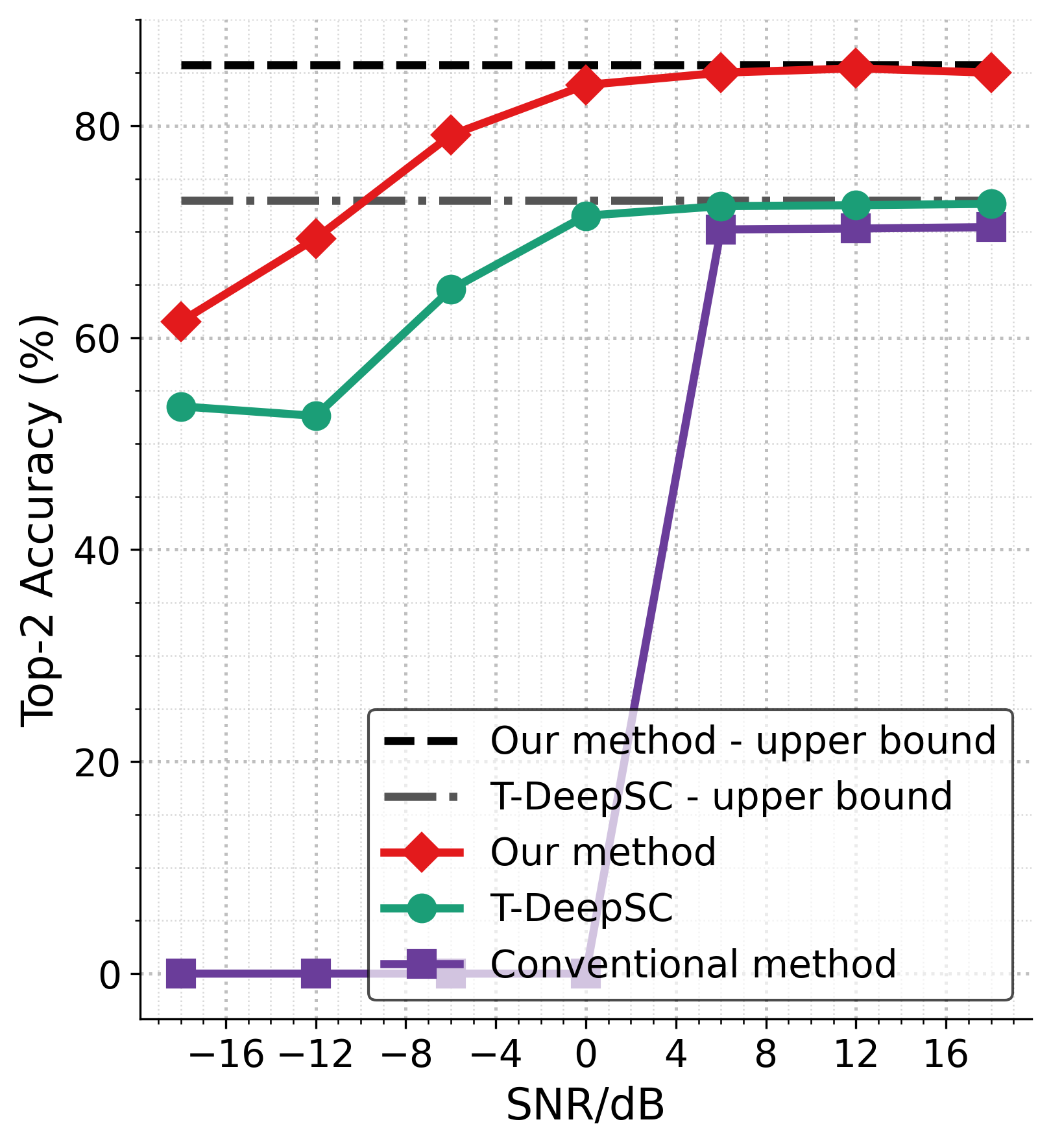}
    \caption*{Top-2 Accuracy in AWGN}
    \label{fig:mosei-awgn-acc2}
  \end{subfigure}\hspace{0.4em}
  \begin{subfigure}[t]{\imgw}
    \centering
    \includegraphics[width=\linewidth]{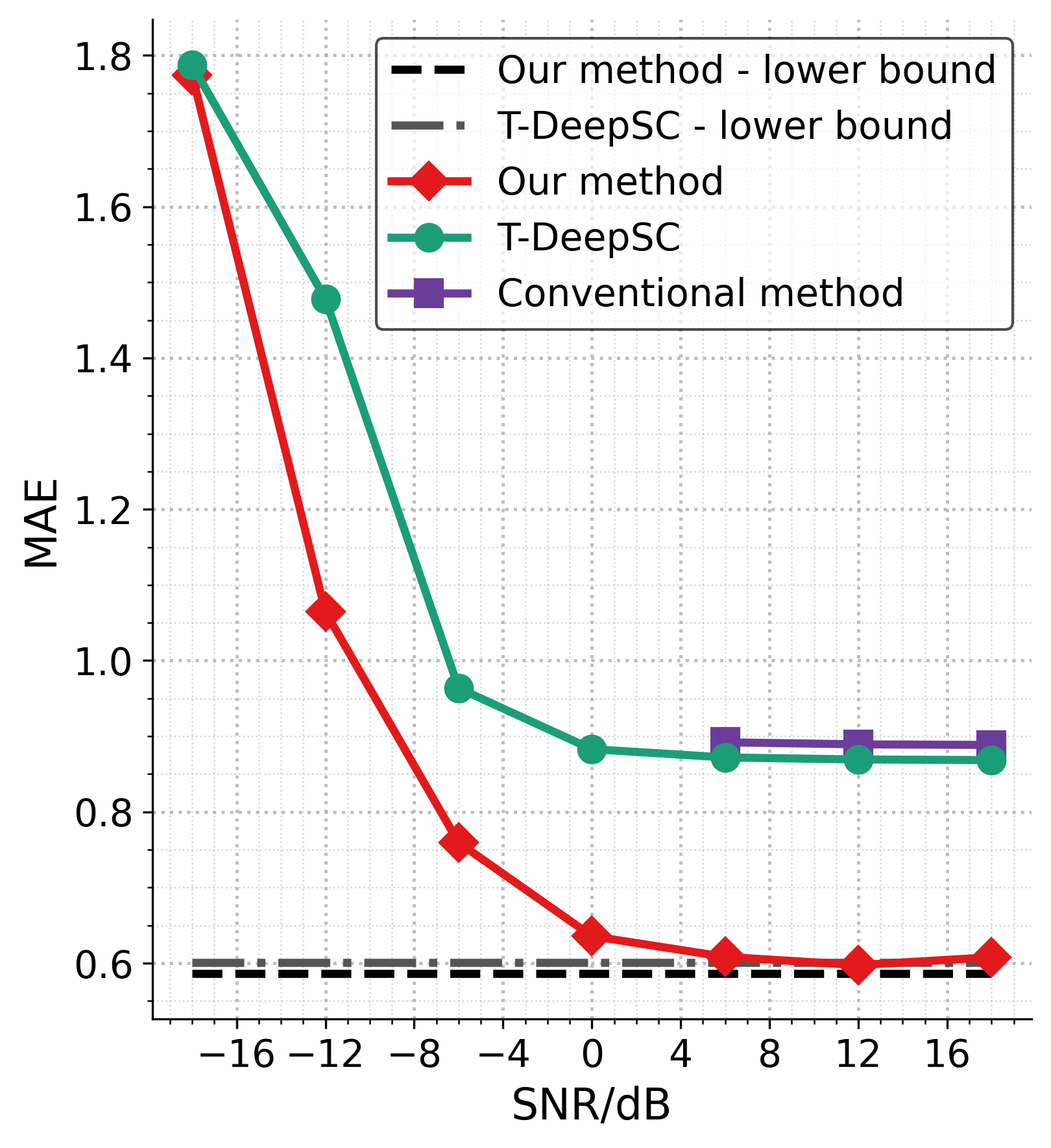}
    \caption*{MAE in AWGN}
    \label{fig:mosei-awgn-mae}
  \end{subfigure}\hspace{0.4em}
  \begin{subfigure}[t]{\imgw}
    \centering
    \includegraphics[width=\linewidth]{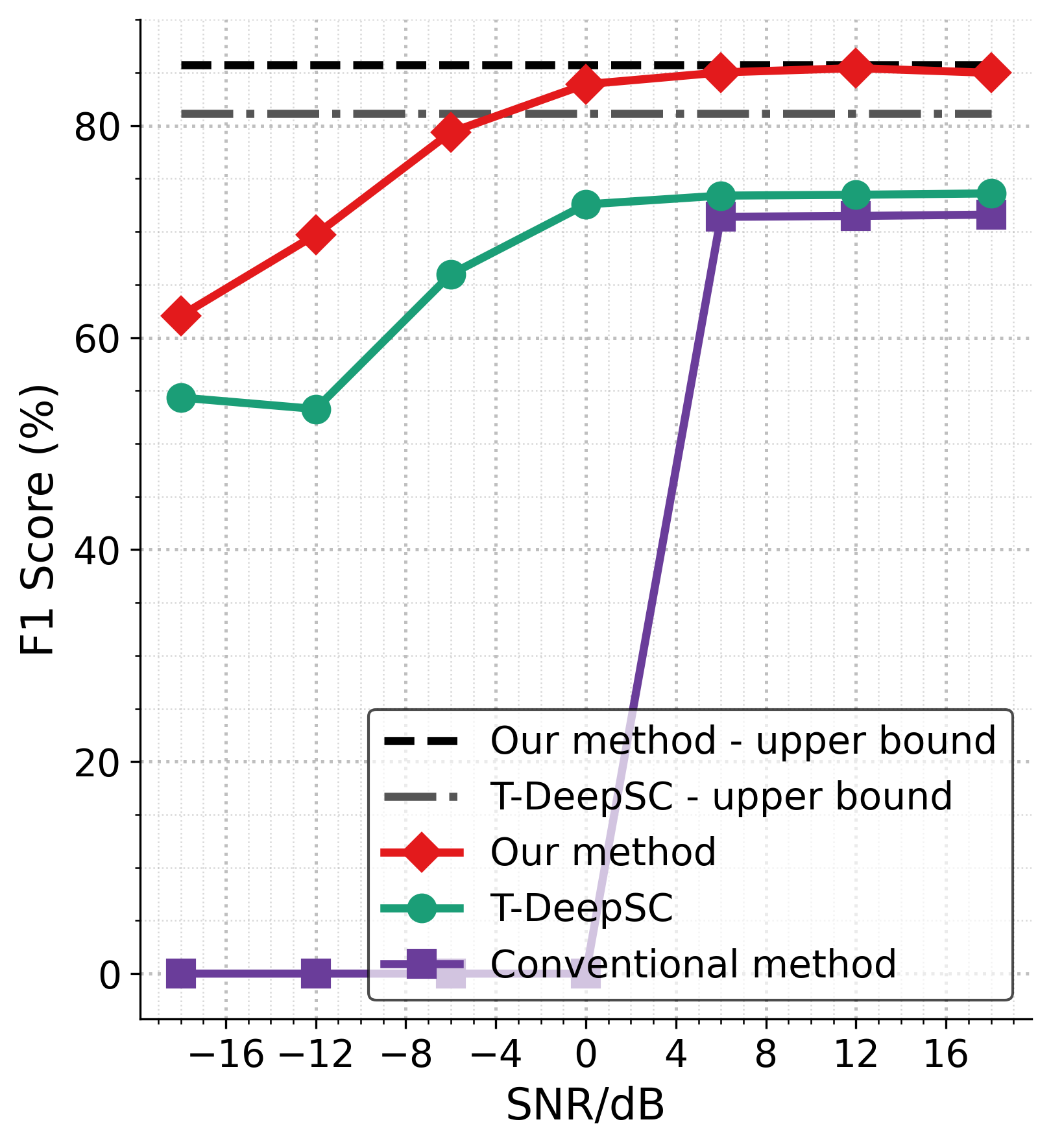}
    \caption*{F1 Score in AWGN}
    \label{fig:mosei-awgn-f1}
  \end{subfigure}\hspace{0.4em}
  \begin{subfigure}[t]{\imgw}
    \centering
    \includegraphics[width=\linewidth]{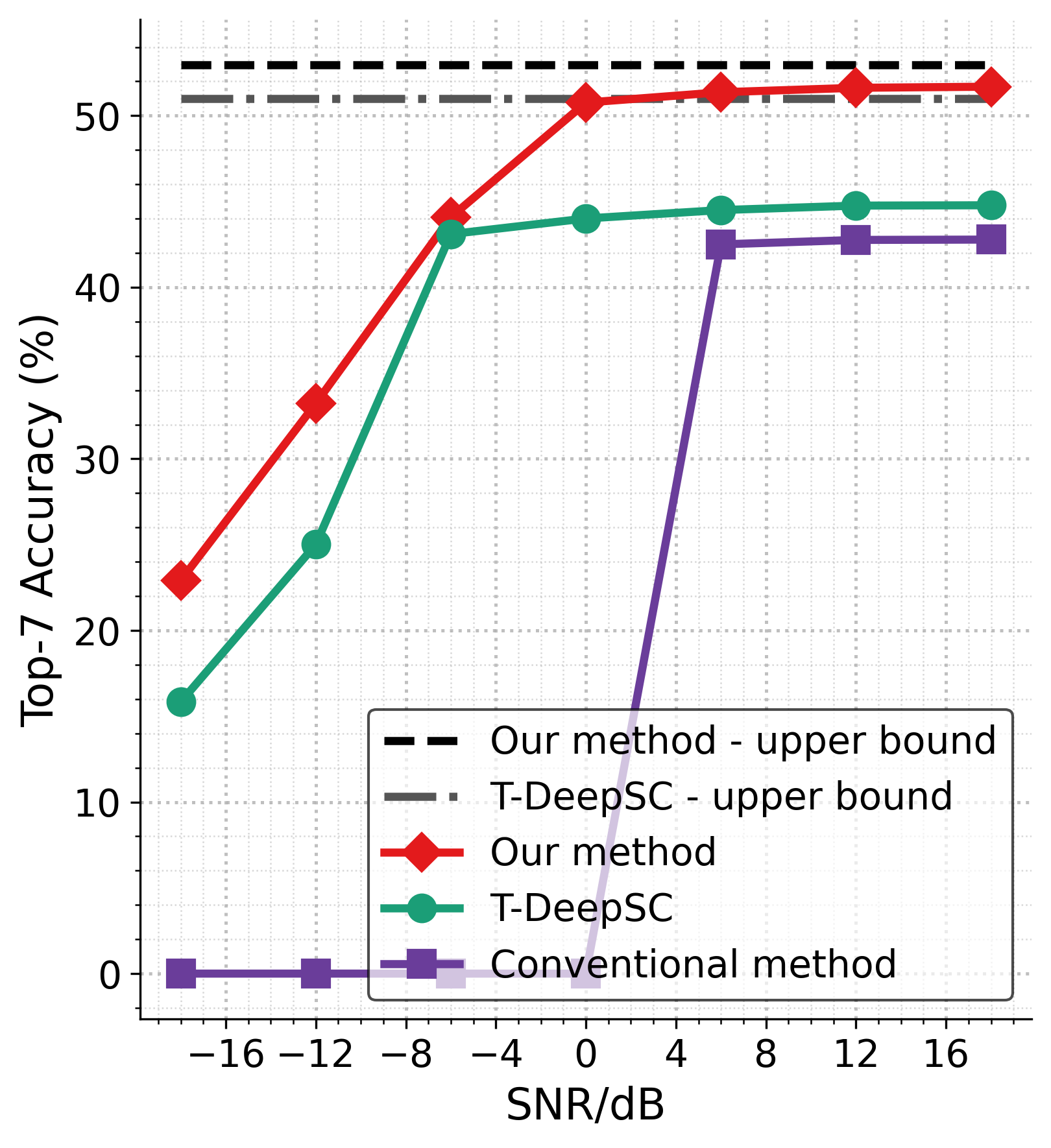}
    \caption*{Top-7 Accuracy in AWGN}
    \label{fig:mosei-awgn-acc7}
  \end{subfigure}

  \vspace{0.8em}

  \begin{subfigure}[t]{\imgw}
    \centering
    \includegraphics[width=\linewidth]{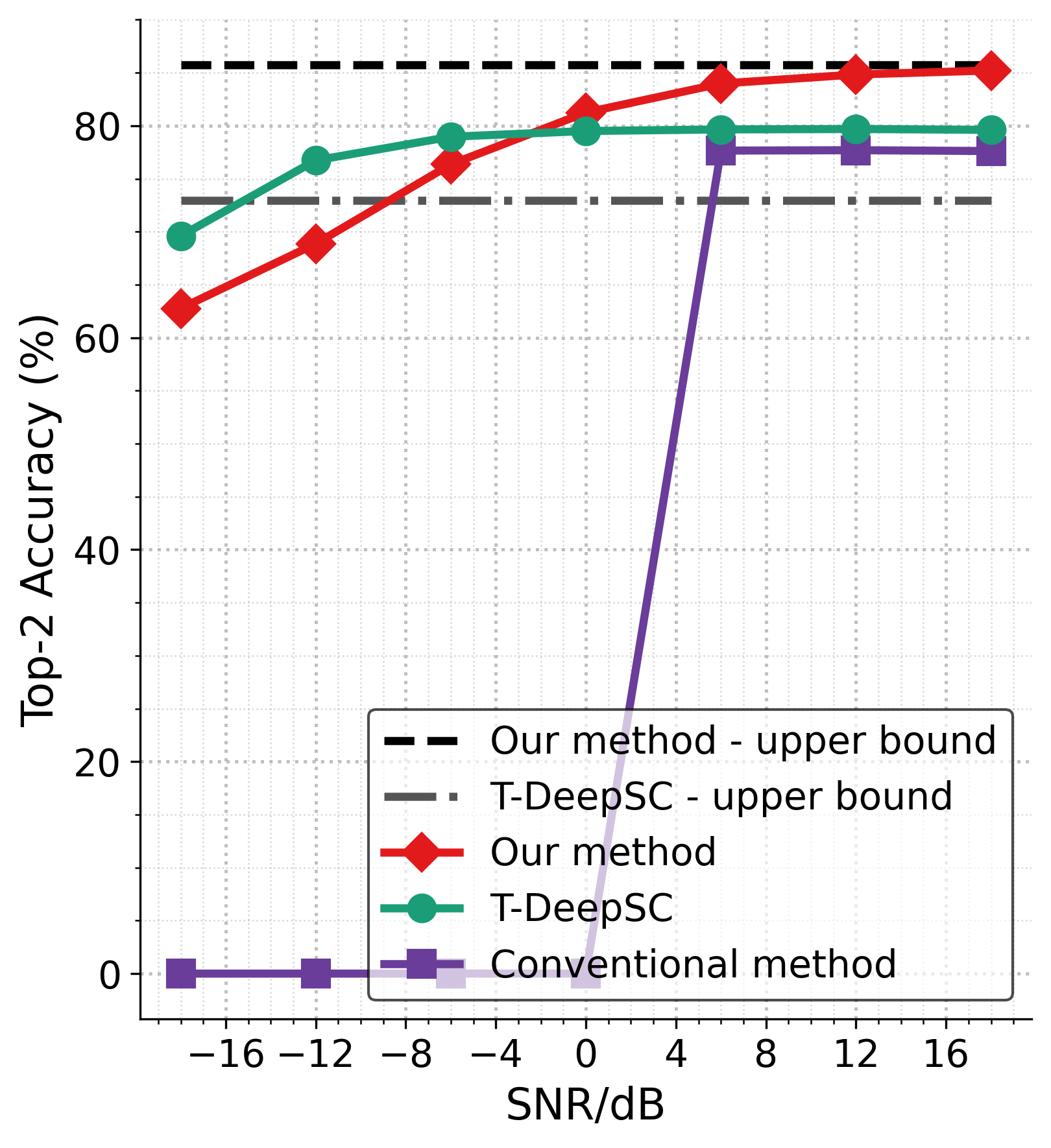}
    \caption*{Top-2 Accuracy in Rayleigh}
    \label{fig:mosei-rayleigh-acc2}
  \end{subfigure}\hspace{0.4em}
  \begin{subfigure}[t]{\imgw}
    \centering
    \includegraphics[width=\linewidth]{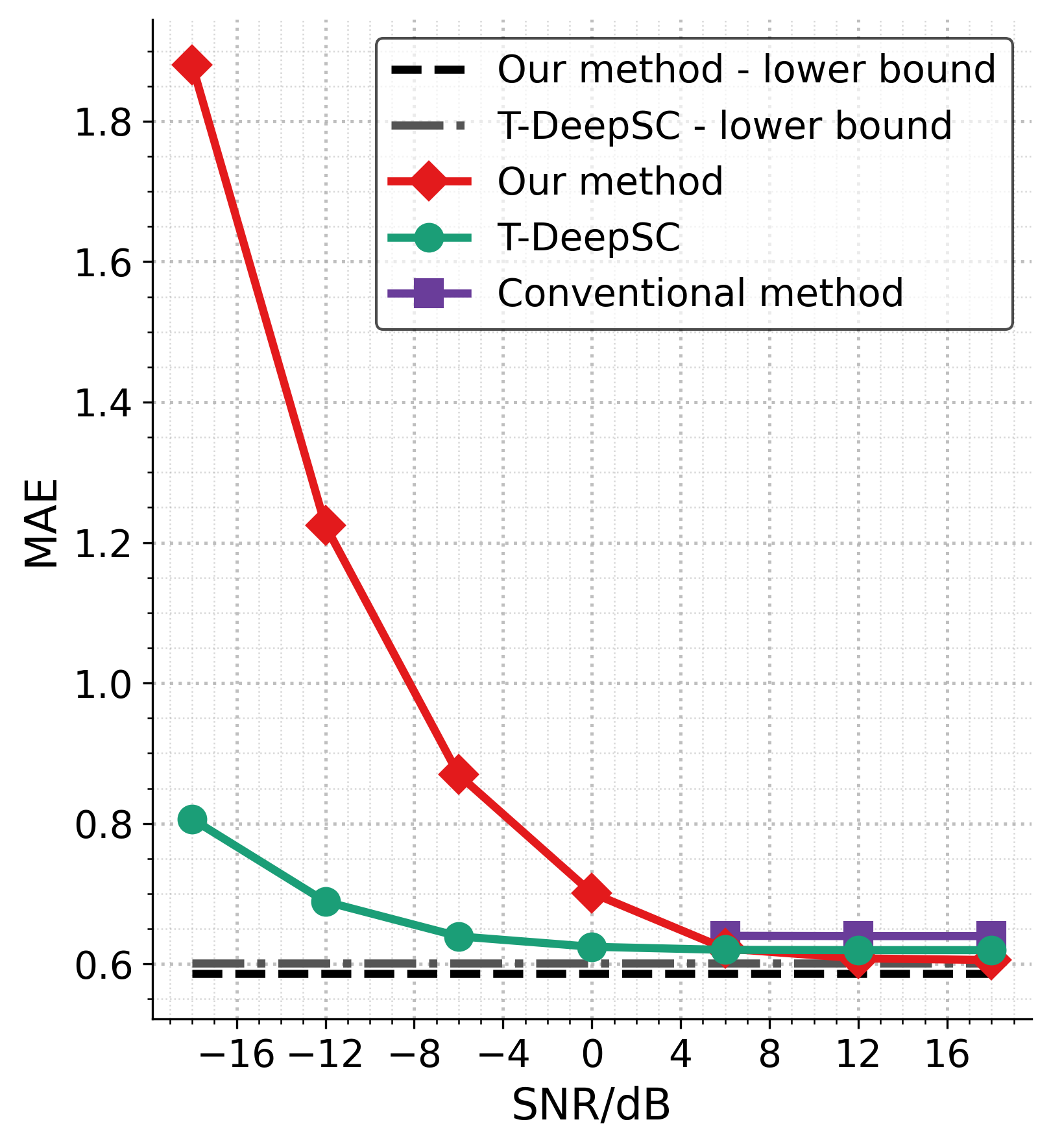}
    \caption*{MAE in Rayleigh}
    \label{fig:mosei-rayleigh-mae}
  \end{subfigure}\hspace{0.4em}
  \begin{subfigure}[t]{\imgw}
    \centering
    \includegraphics[width=\linewidth]{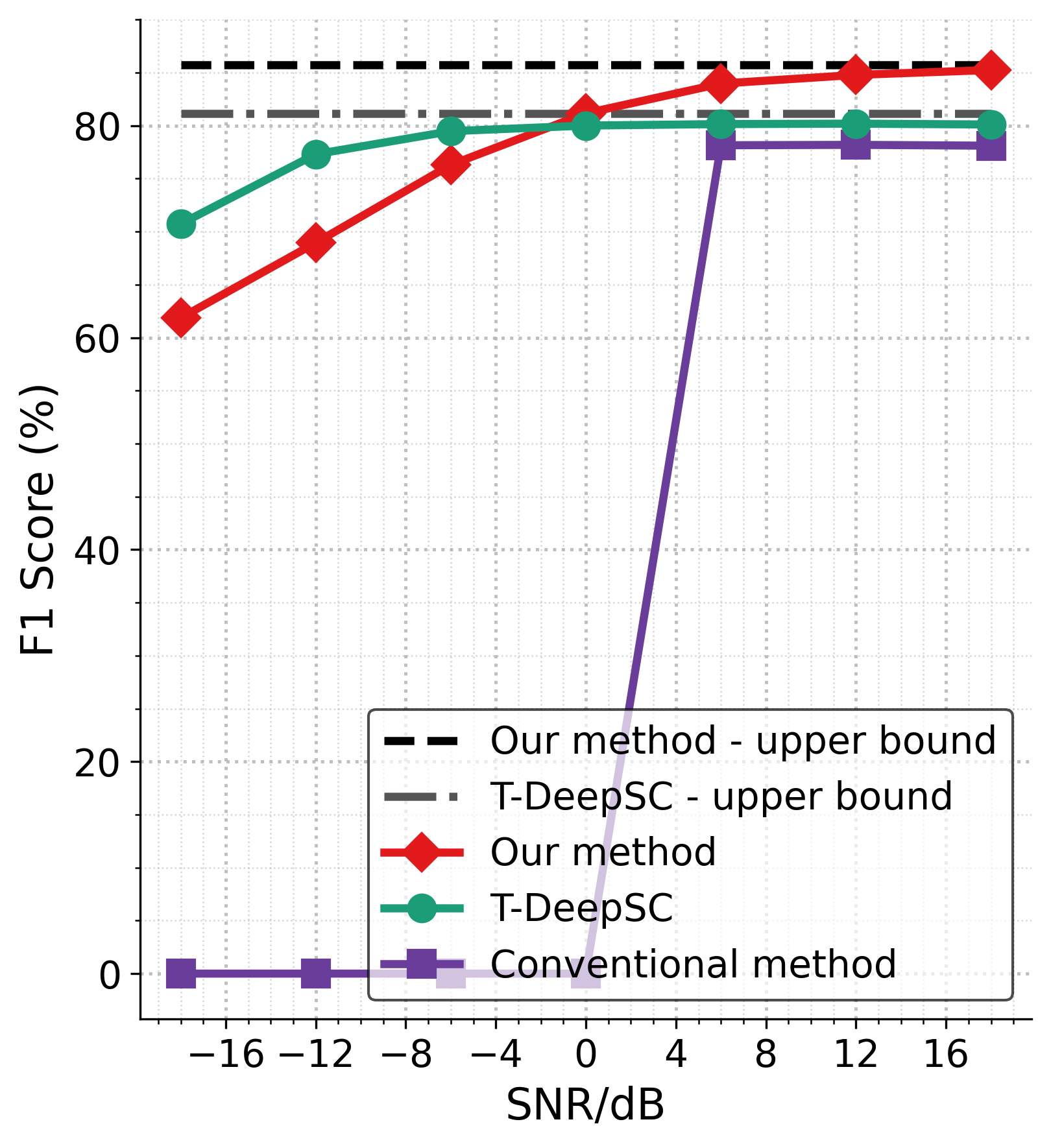}
    \caption*{F1 Score in Rayleigh}
    \label{fig:mosei-rayleigh-f1}
  \end{subfigure}\hspace{0.4em}
  \begin{subfigure}[t]{\imgw}
    \centering
    \includegraphics[width=\linewidth]{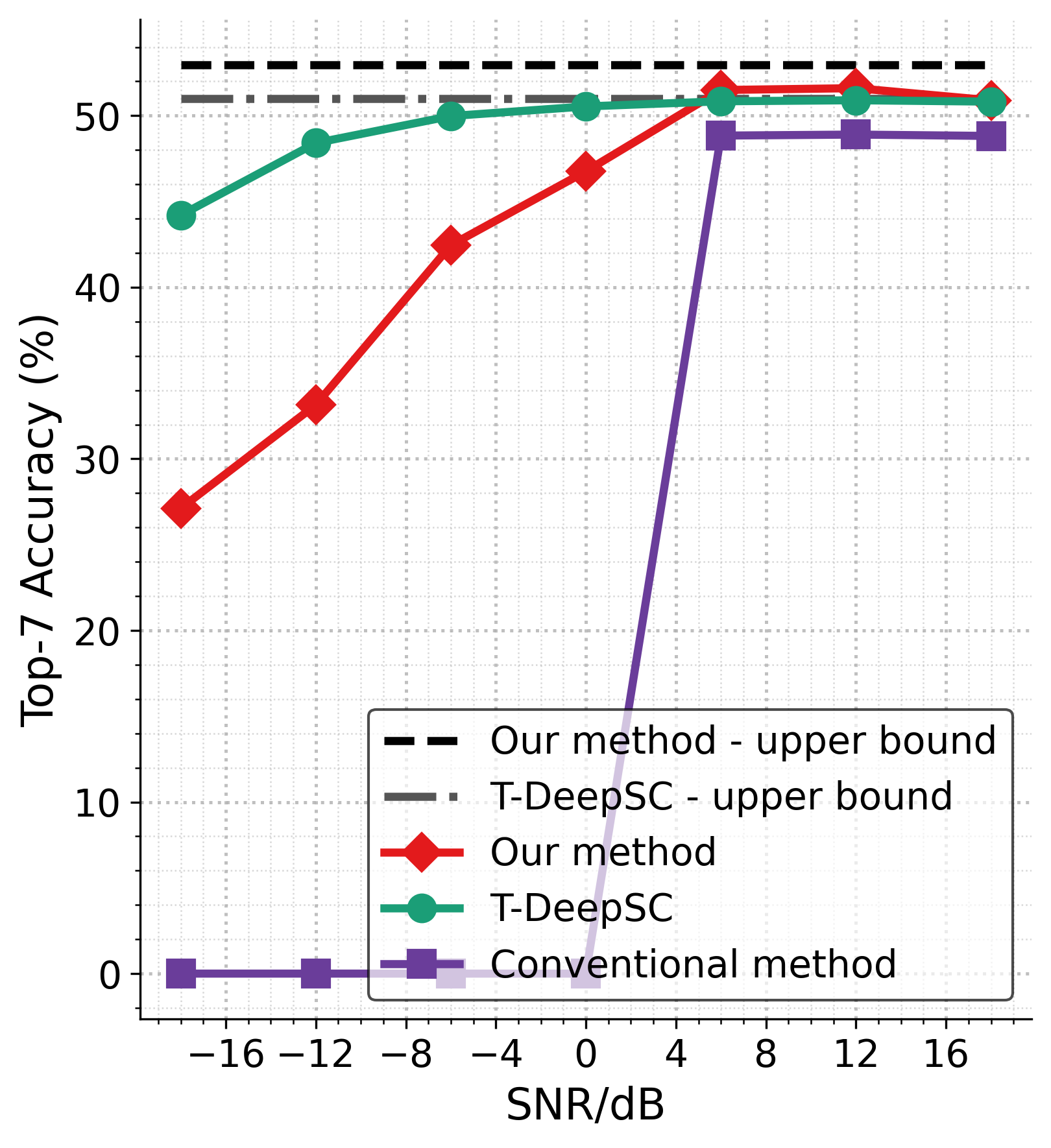}
    \caption*{Top-7 Accuracy in Rayleigh}
    \label{fig:mosei-rayleigh-acc7}
  \end{subfigure}

\caption{System performance on the MOSEI dataset under AWGN and Rayleigh fading channels.}

  \label{fig:mosei-2x4}
\end{figure*}
\subsection{Overall Objective Function}

For the whole system, the objective function is 
\begin{align}
\mathcal{L}=\sum\nolimits_{m}\mathcal{L}_{\text{U-VIB}}+\mathcal{L}_{\text{M-VIB}}+\lambda_{\text{red}}\mathcal{L}_{\text{red}},
\label{eq:totalLoss} 
\end{align}
where $\lambda_{\text{red}}\ge 0$ is a hyperparameter that controls the trade-off  between the redundancy reduction loss term and the VIB losses.  $\mathcal{L}$ comprises three components: (i) the uni-modal VIB loss $\mathcal{L}_{\text{U-VIB}}$ calculated by \eqref{eq:loss_uvib}, which applies VIB to each individual modality; (ii) the multi-modal VIB loss $\mathcal{L}_{\text{M-VIB}}$ calculated by \eqref{eq:loss_mvib}, which further compresses the fused representation to reduce residual inter-modal redundancy and improve robustness against channel noise; and (iii) the redundancy reduction loss $\mathcal{L}_{\text{red}}$ calculated by \eqref{eq:loss_red}, which uses adversarial training to minimize the MI between modality pairs, encouraging complementary rather than redundancy.




\subsection{One-step Training}

To train the whole encoder/decoder NN, we adopt a one-step training strategy to jointly train all modules in a single forward–backward pass, as shown in Algorithm~\ref{alg:cmib-train}. At each iteration, modality-specific encoders extract features, which are then processed by the U-VIB modules. The fused representation is then refined via cross-modality MI reduction with GRL adversarial training, and then transmitted through wireless channels and decoded by the M-VIB module. All parameters are updated simultaneously, improving both efficiency and stability.

\begin{figure*}[t]
  \centering
  \newcommand{\imgw}{0.23\linewidth}

  \begin{subfigure}[t]{\imgw}
    \centering
    \includegraphics[width=\linewidth]{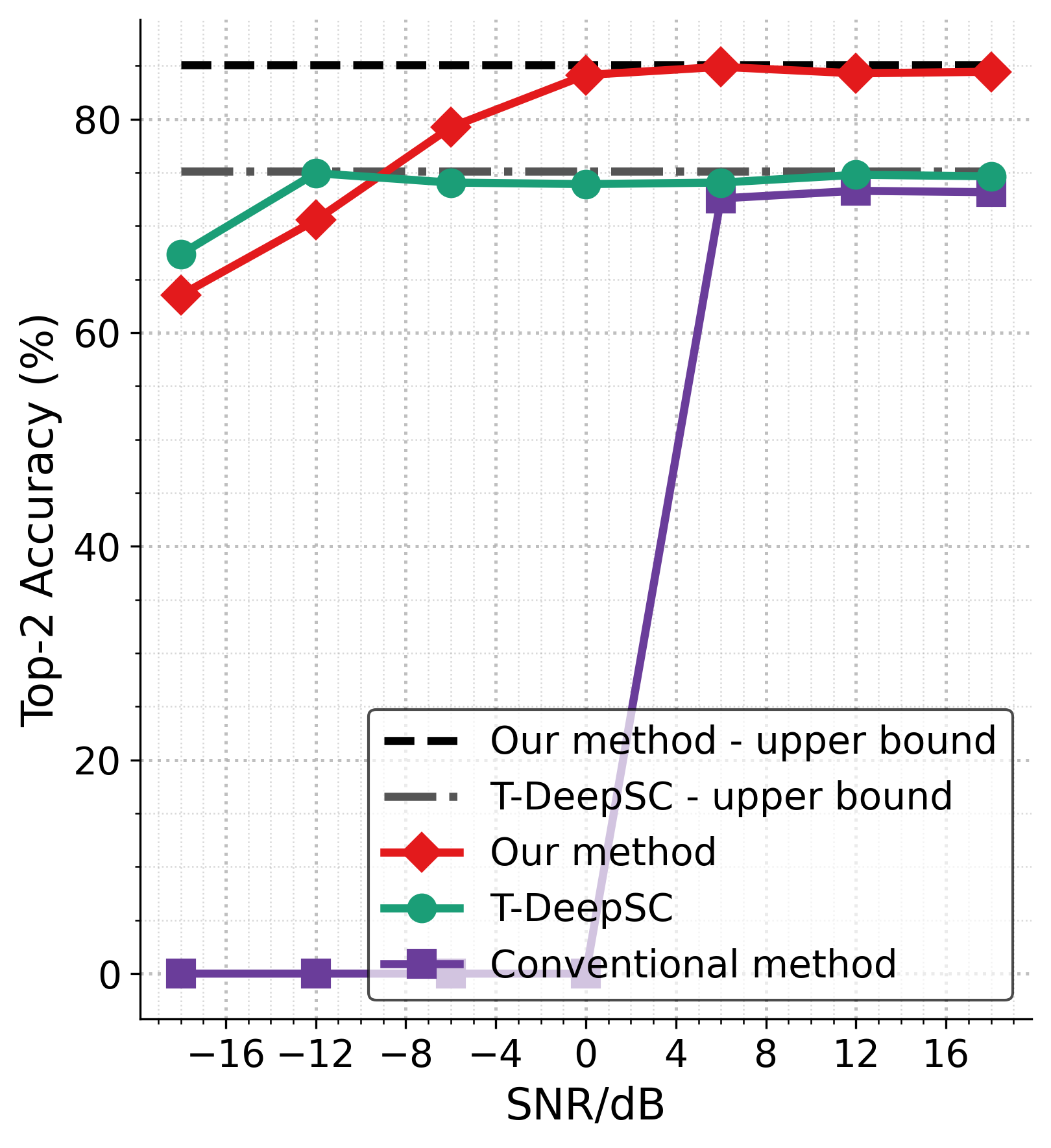}
    \caption*{Top-2 Accuracy in AWGN}
    \label{fig:mosi-awgn-acc2}
  \end{subfigure}\hspace{0.4em}
  \begin{subfigure}[t]{\imgw}
    \centering
    \includegraphics[width=\linewidth]{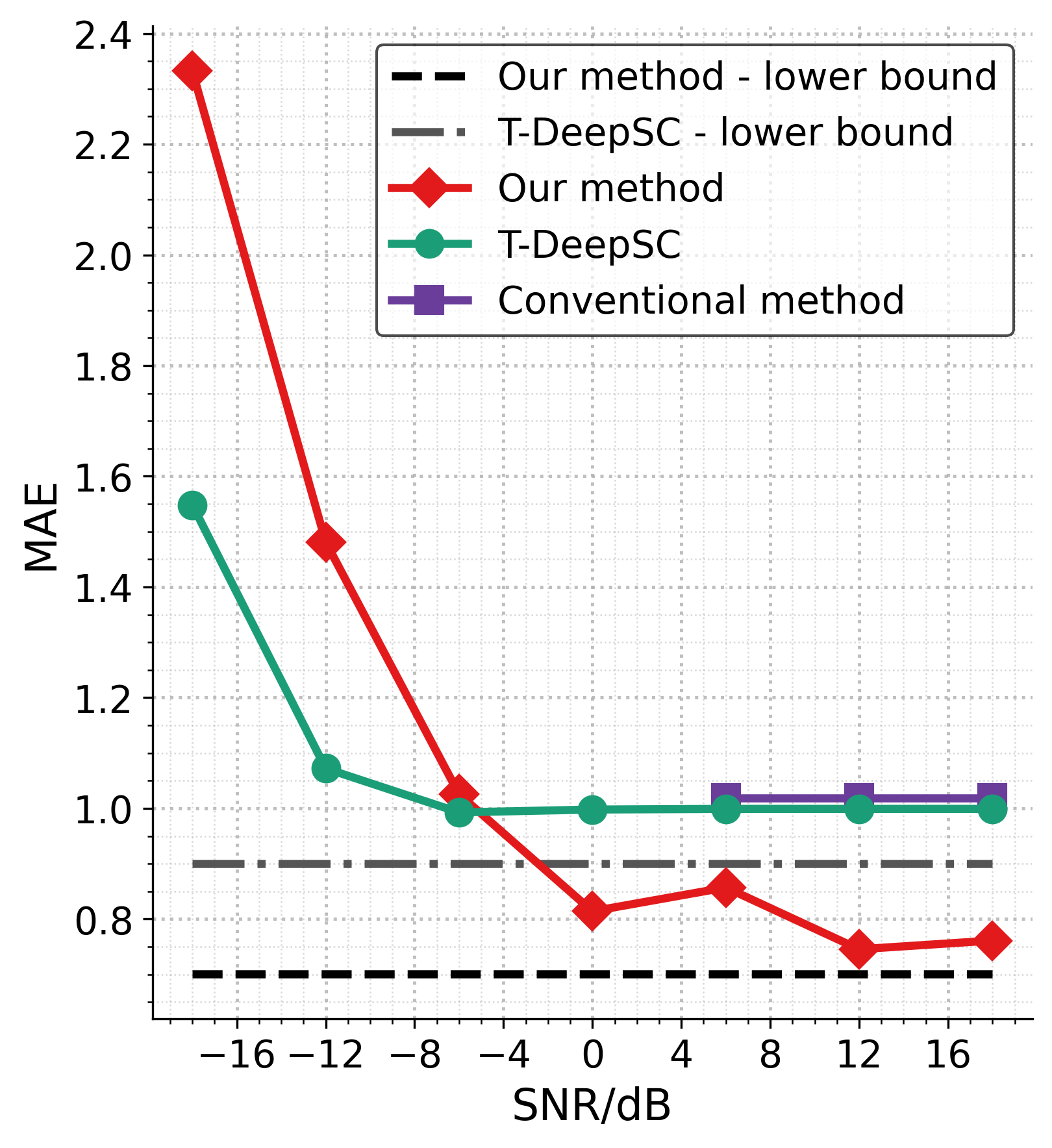}
    \caption*{MAE in AWGN}
    \label{fig:mosi-awgn-mae}
  \end{subfigure}\hspace{0.4em}
  \begin{subfigure}[t]{\imgw}
    \centering
    \includegraphics[width=\linewidth]{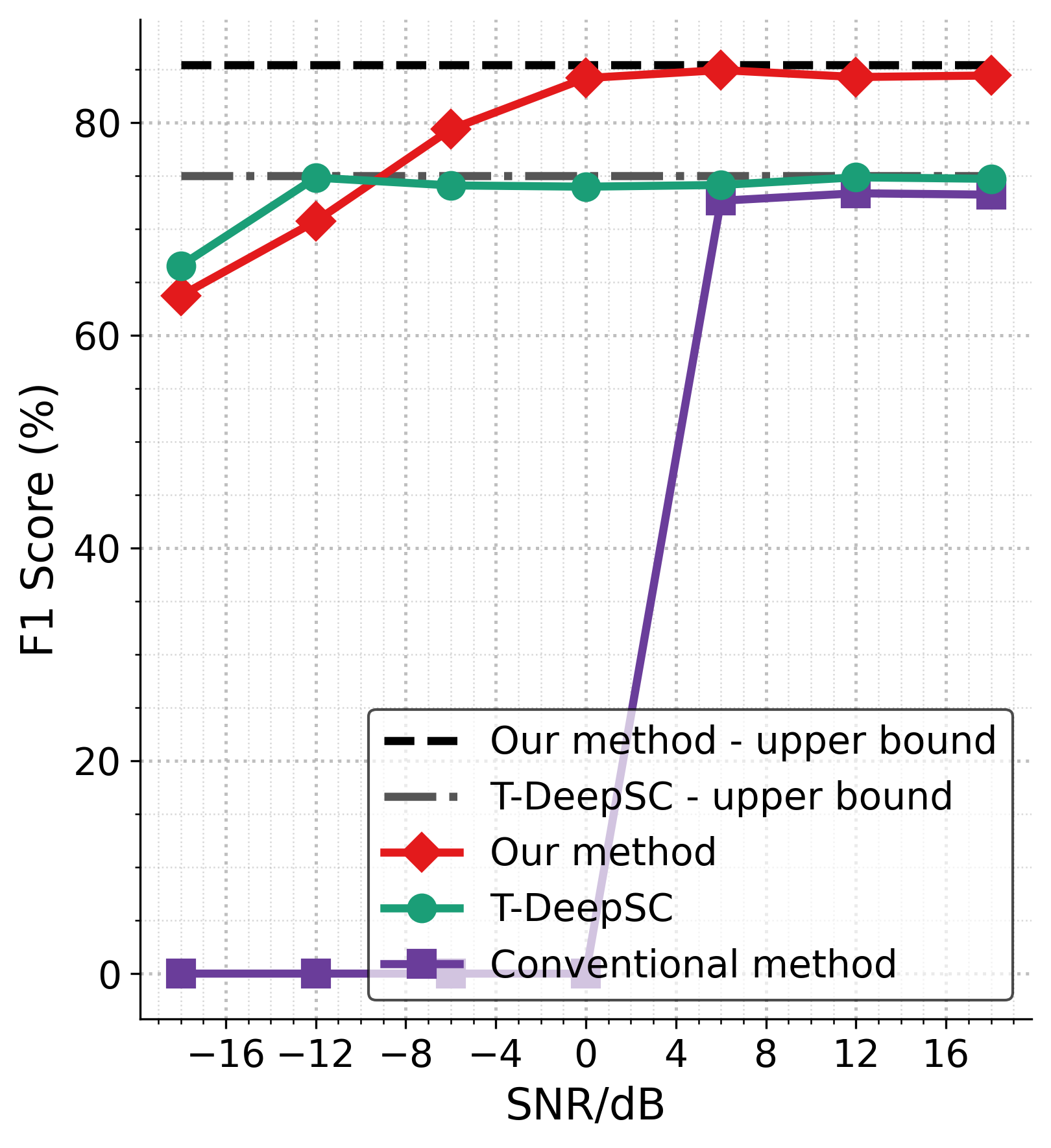}
    \caption*{F1 Score in AWGN}
    \label{fig:mosi-awgn-f1}
  \end{subfigure}\hspace{0.4em}
  \begin{subfigure}[t]{\imgw}
    \centering
    \includegraphics[width=\linewidth]{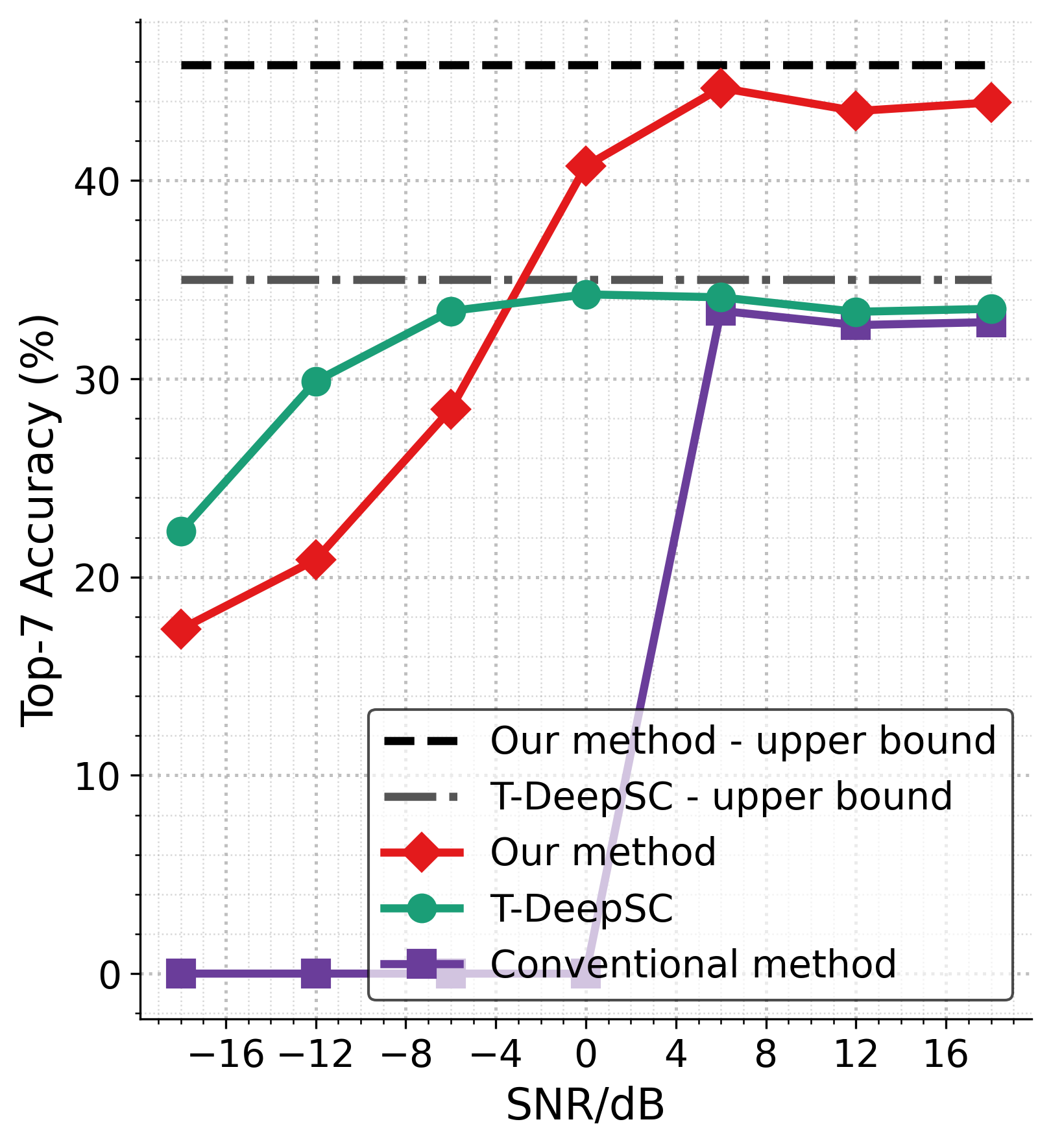}
    \caption*{Top-7 Accuracy in AWGN}
    \label{fig:mosi-awgn-acc7}
  \end{subfigure}

  \vspace{0.8em}

  \begin{subfigure}[t]{\imgw}
    \centering
    \includegraphics[width=\linewidth]{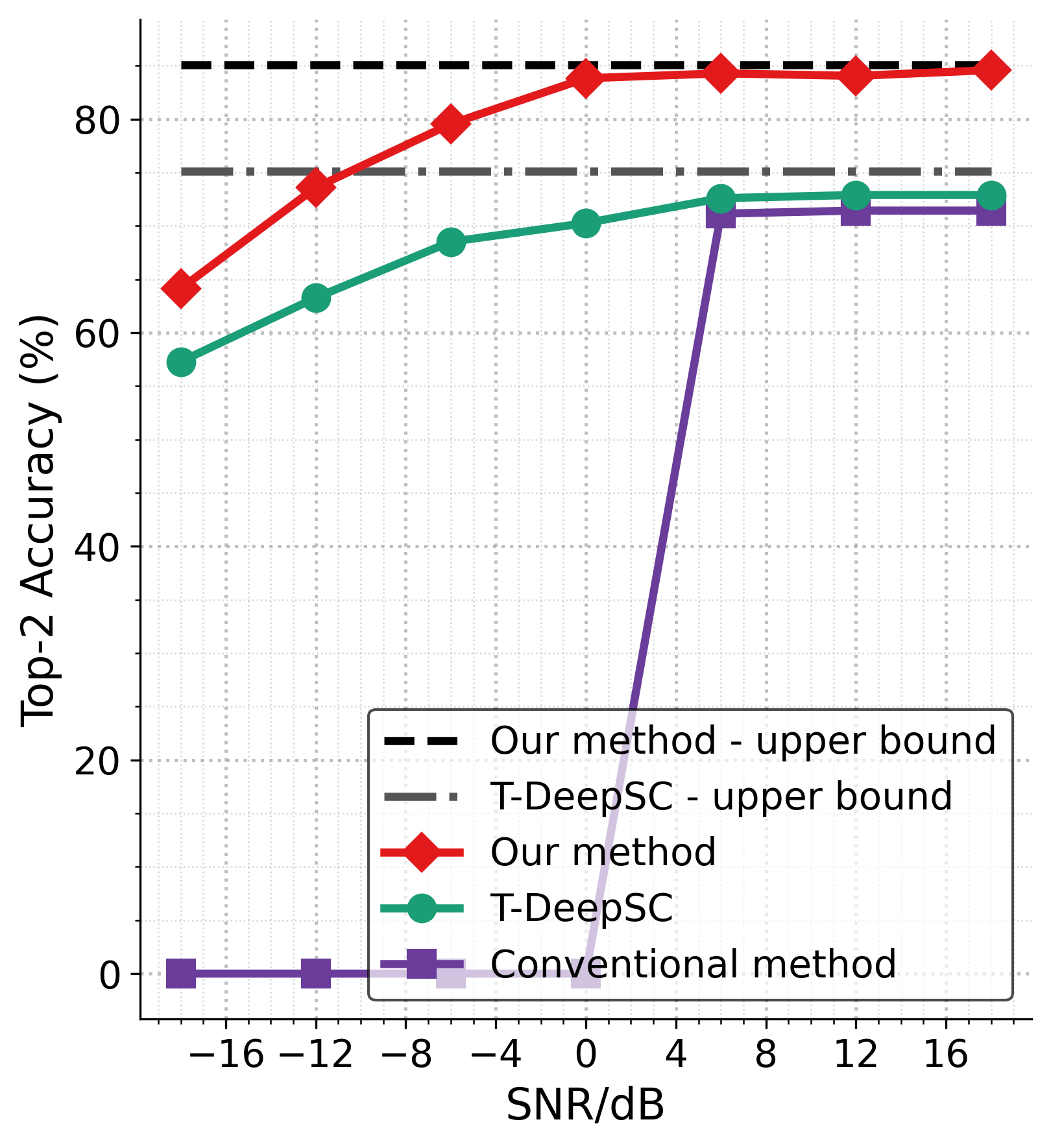}
    \caption*{Top-2 Accuracy in Rayleigh}
    \label{fig:mosi-rayleigh-acc2}
  \end{subfigure}\hspace{0.4em}
  \begin{subfigure}[t]{\imgw}
    \centering
    \includegraphics[width=\linewidth]{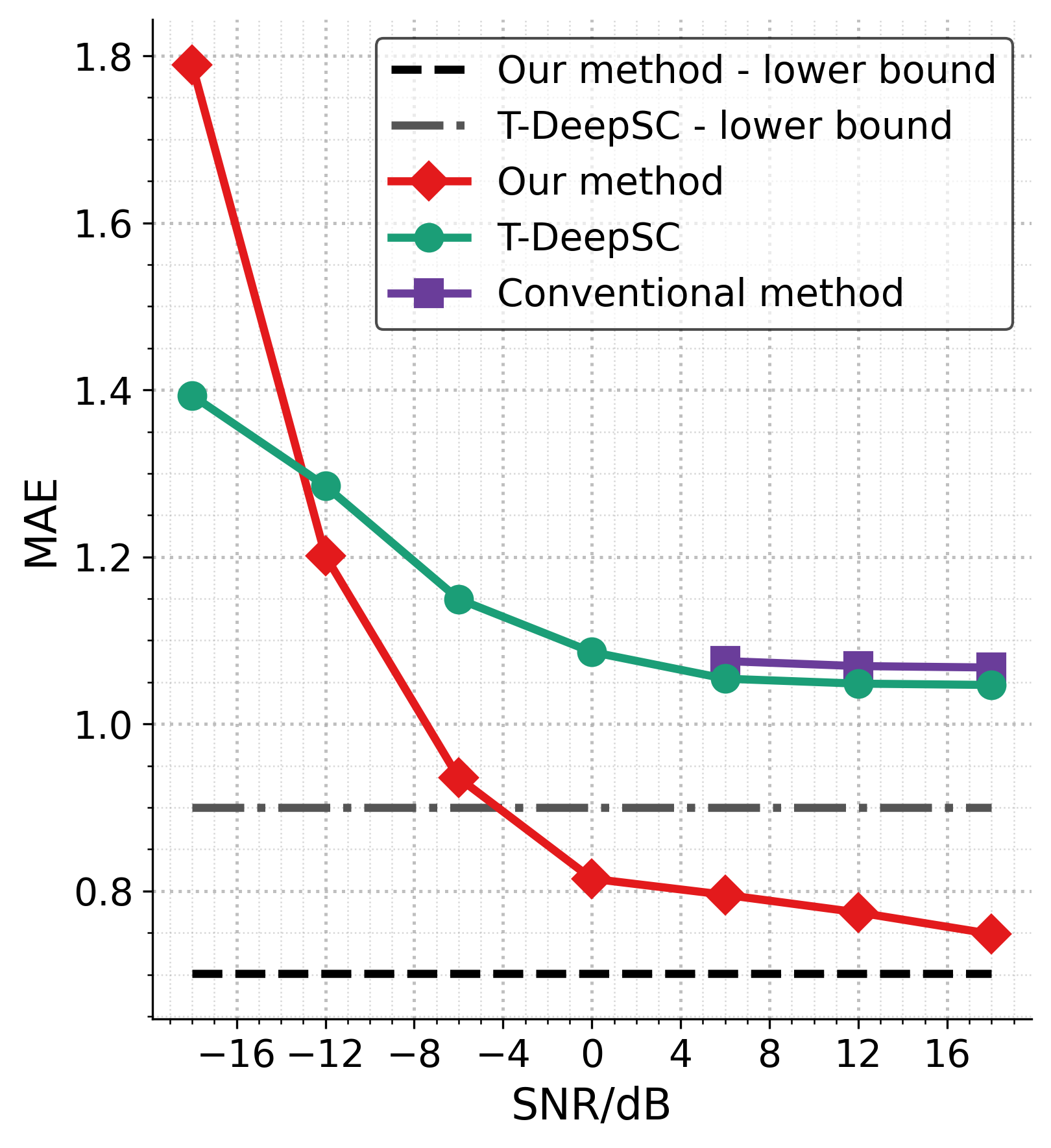}
    \caption*{MAE in Rayleigh}
    \label{fig:mosi-rayleigh-mae}
  \end{subfigure}\hspace{0.4em}
  \begin{subfigure}[t]{\imgw}
    \centering
    \includegraphics[width=\linewidth]{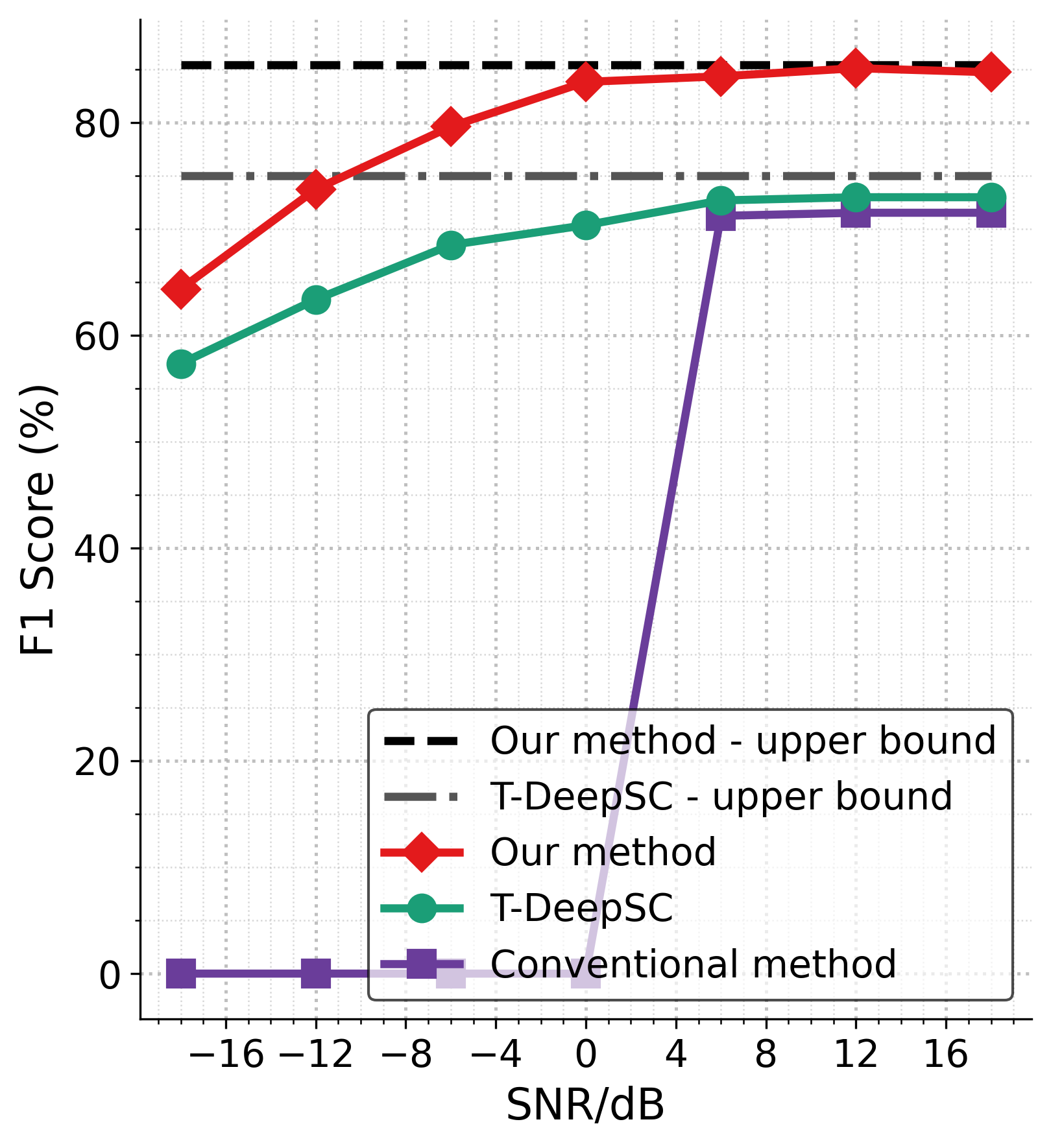}
    \caption*{F1 Score in Rayleigh}
    \label{fig:mosi-rayleigh-f1}
  \end{subfigure}\hspace{0.4em}
  \begin{subfigure}[t]{\imgw}
    \centering
    \includegraphics[width=\linewidth]{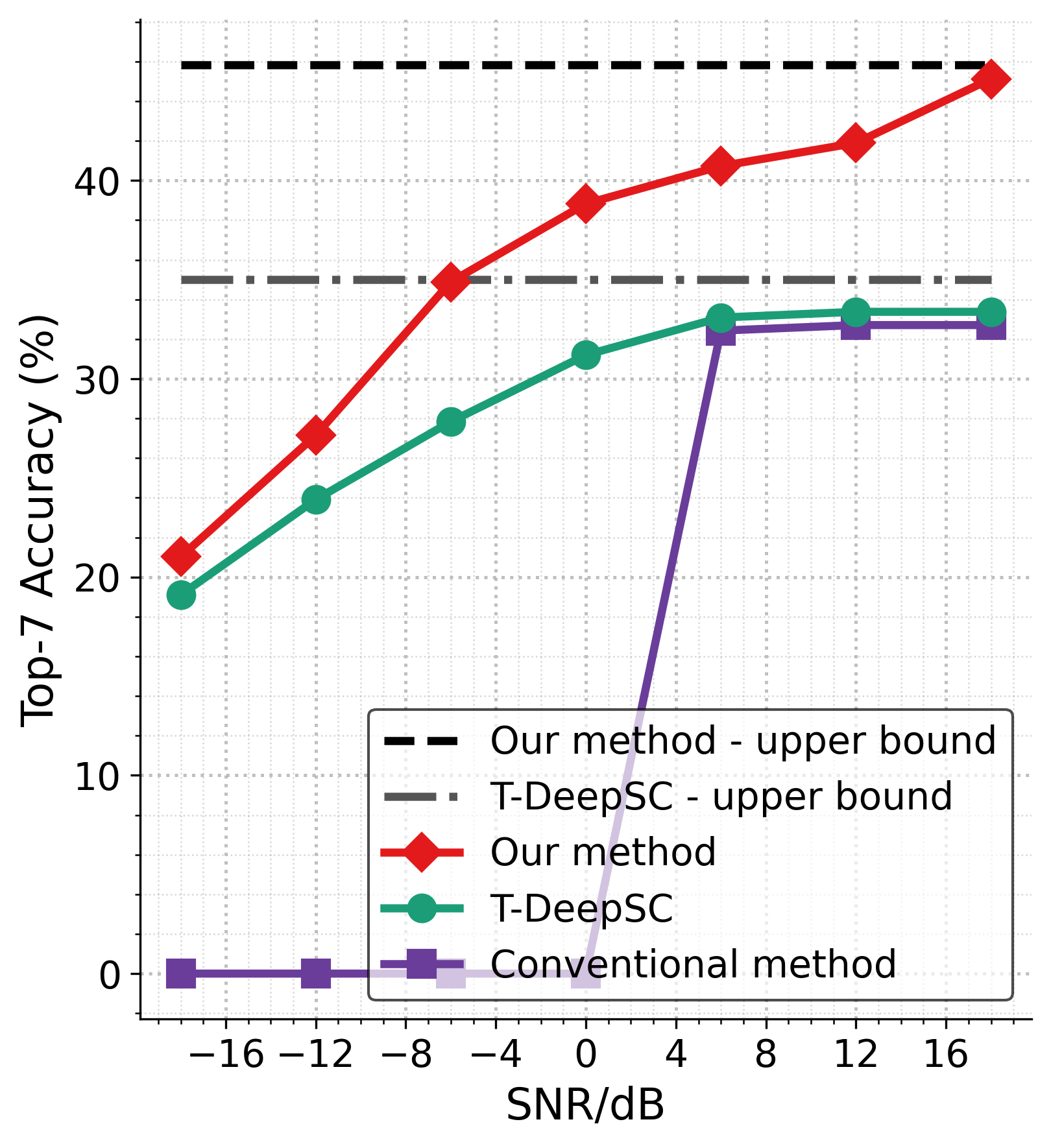}
    \caption*{Top-7 Accuracy in Rayleigh}
    \label{fig:mosi-rayleigh-acc7}
  \end{subfigure}

\caption{System performance on the MOSI dataset under AWGN and Rayleigh fading channel.}

  \label{fig:mosi-2x4}
\end{figure*}

\section{Numerical Results}
\label{sec:experiment}

\subsection{Experimental Setups}
\subsubsection{Datasets}
We use two datasets in experiments. Specifically, CMU-MOSI \cite{mosi} is a multi-modal dataset for sentiment analysis tasks. It consists of  $93$ opinion video clips and is further divided into $2,199$ short video clips. The clips are annotated in a range $[-3,3]$, where $-3$ indicates the strongest negative sentiment and $+3$ is the strongest positive sentiment. CMU-MOSEI \cite{mosei} is a big multi-modal dataset for sentiment analysis tasks with 2928 videos annotated in a range of $[-3,3]$.

\subsubsection{Baselines}
We compare the proposed method with the following baselines:
\begin{itemize}
    \item T-DeepSC \cite{multi_modal_task}: A unified multi-task semantic communication framework designed for multi-modal data. T-DeepSC jointly optimizes source and channel coding within a deep learning architecture, enabling TOC for multiple modalities.
    \item Conventional video compression: We use H.264 standard \cite{wiegand2003overview} for video sentiment feature transmission over wireless channels.
    \item Upper bound / Lower bound: Performance bounds for transmitting features over noiseless channels.
\end{itemize}

\subsubsection{Performance metrics}






The performance is evaluated using four metrics: Top-7 accuracy, Top-2 accuracy, F1, and mean absolute error (MAE) scores. Specifically, Top-7 and Top-2 accuracies denote the classification accuracy over 7-class and binary sentiment (positive or negative) categories, respectively. F1 score represents the harmonic mean of the precision and recall, providing a balanced measure for imbalanced classification. MAE score quantifies the average absolute deviation between predicted and ground-truth continuous scores. Higher values are better for Top-7 accuracy, Top-2 accuracy, and F1, whereas lower values are preferred for MAE.

\subsubsection{Implementation}

Following the standard pre-processing procedure of the CMU multi-modal datasets, we obtain 768-dimensional text embeddings via the BERT-base-uncased model.  For visual features, CMU-MOSEI dataset yields 47-dimensional representations extracted with the collaborative voice analysis repository for speech technologies (COVAREP) \cite{degottex2014covarep}, while CMU-MOSI dataset yields 35-dimensional representations extracted with the facial action coding environment (FACET) based on the facial action coding system (FACS). Acoustic features of 74 dimensions are also obtained using COVAREP. The image and text transmitters are initialized with pre-trained transformer and BERT, respectively. Simulations are conducted over AWGN and Rayleigh fading channels. During training, the batch size is fixed to 32, and the network is optimized with 50 epochs. In (\ref{eq:totalLoss}), $\lambda_{\text{red}}$ is set to 0.4. To further enhance the stability of adversarial optimization, we adopt warm-up strategies: the gradient reversal factor $\alpha$ in (\ref{eq:grl_alpha}) is linearly increased from 0 to 1 over the first warm epochs $E_{\text{warm}}=3$, while $\lambda_{\text{red}}$ is set to 0 during the initial $E_{\text{warm}}$ epochs and then linearly increases to 0.4 during the subsequent epochs. We train our model at random SNRs from 0 to 21 dB. 

\subsection{Results Analysis}
\subsubsection{Task performance}

The results on MOSEI and MOSI datasets are shown in Fig. \ref{fig:mosei-2x4} and Fig. \ref{fig:mosi-2x4}, respectively. For both AWGN and Rayleigh fading channels, our proposed model outperforms the two baselines.
In AWGN channels, our method achieves consistently higher classification accuracies (Top-2, F1 Score, and Top-7) across all SNRs, approaching the upper bound even in low SNR regimes. For instance, for MOSEI dataset with $SNR=-6$ dB, our model achieves an F1 score of $79.38\% $ and a Top-7 Accuracy of $44.09\%$, yielding $13.40\%$ and $0.99\%$ improvement over T-DeepSC ($65.98\%$ and $43.10\%)$, respectively. In contrast, the conventional method fails to operate reliably under such SNR. Moreover, it yields lower MAE compared to both baselines, indicating better robustness to various channel conditions. In Rayleigh fading channels,  our method maintains stable performance under low-SNR regimes, while T-DeepSC and the conventional methods experience significant degradation. For example, on MOSI dataset at $-12$~dB under Rayleigh fading channel, our model achieves a Top-2 accuracy of $73.59\%$ and an F1 score of $73.71\%$, surpassing T-DeepSC ($63.27\%$ and $63.35\%$) by $10.32\%$ and $10.40\%$, respectively.
The results show that the proposed method not only improves semantic transmission accuracy but also enhances resilience in wireless channels.

\subsubsection{Effect of the adversarial redundancy reduction}
\begin{figure*}[t]
    \centering
    \begin{subfigure}[t]{0.32\textwidth}
        \centering
        \includegraphics[width=\textwidth]{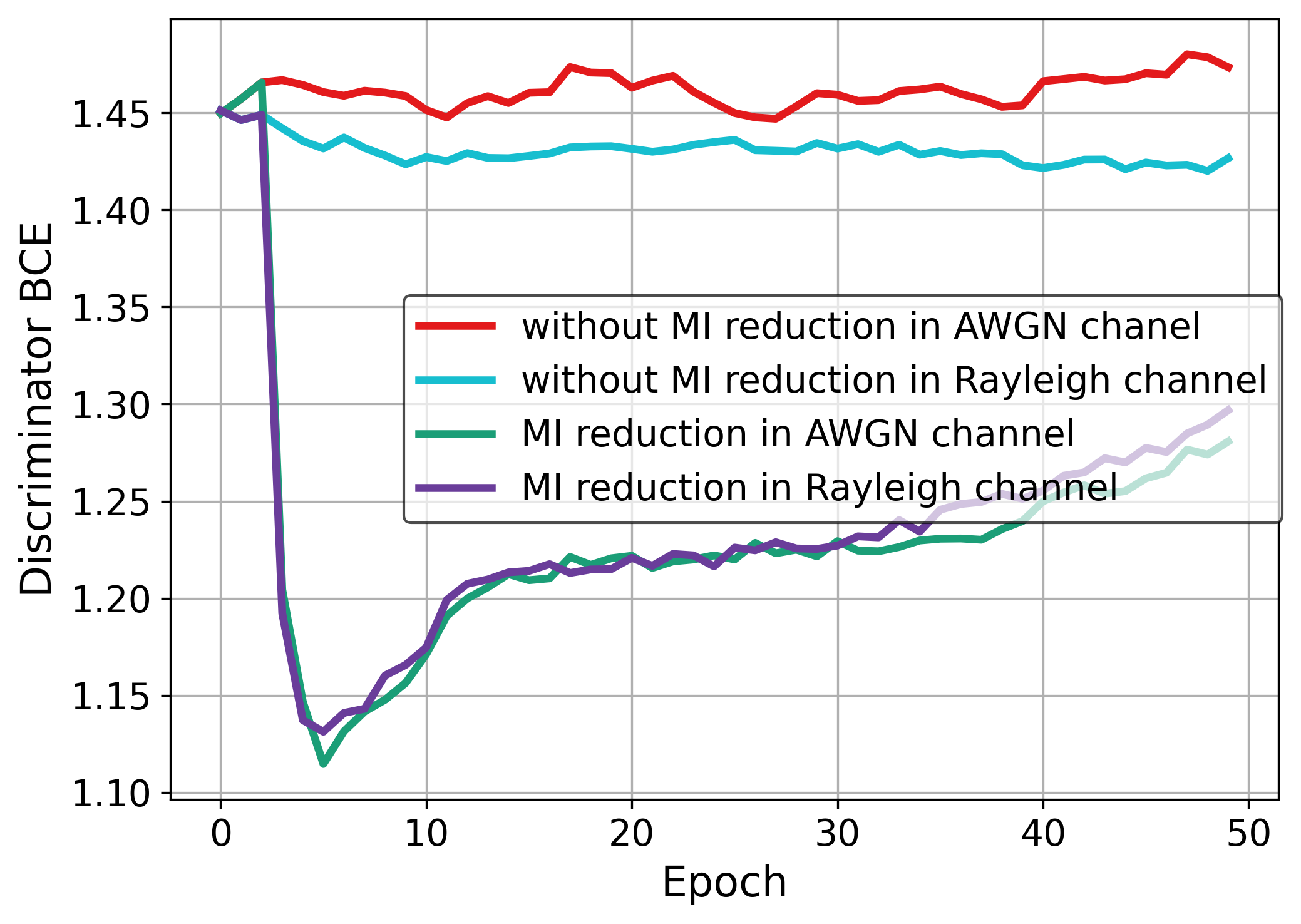}
        \subcaption{BCE between image and audio.}
        \label{fig:mi_va}
    \end{subfigure}%
    \hfill
    \begin{subfigure}[t]{0.32\textwidth}
        \centering
        \includegraphics[width=\textwidth]{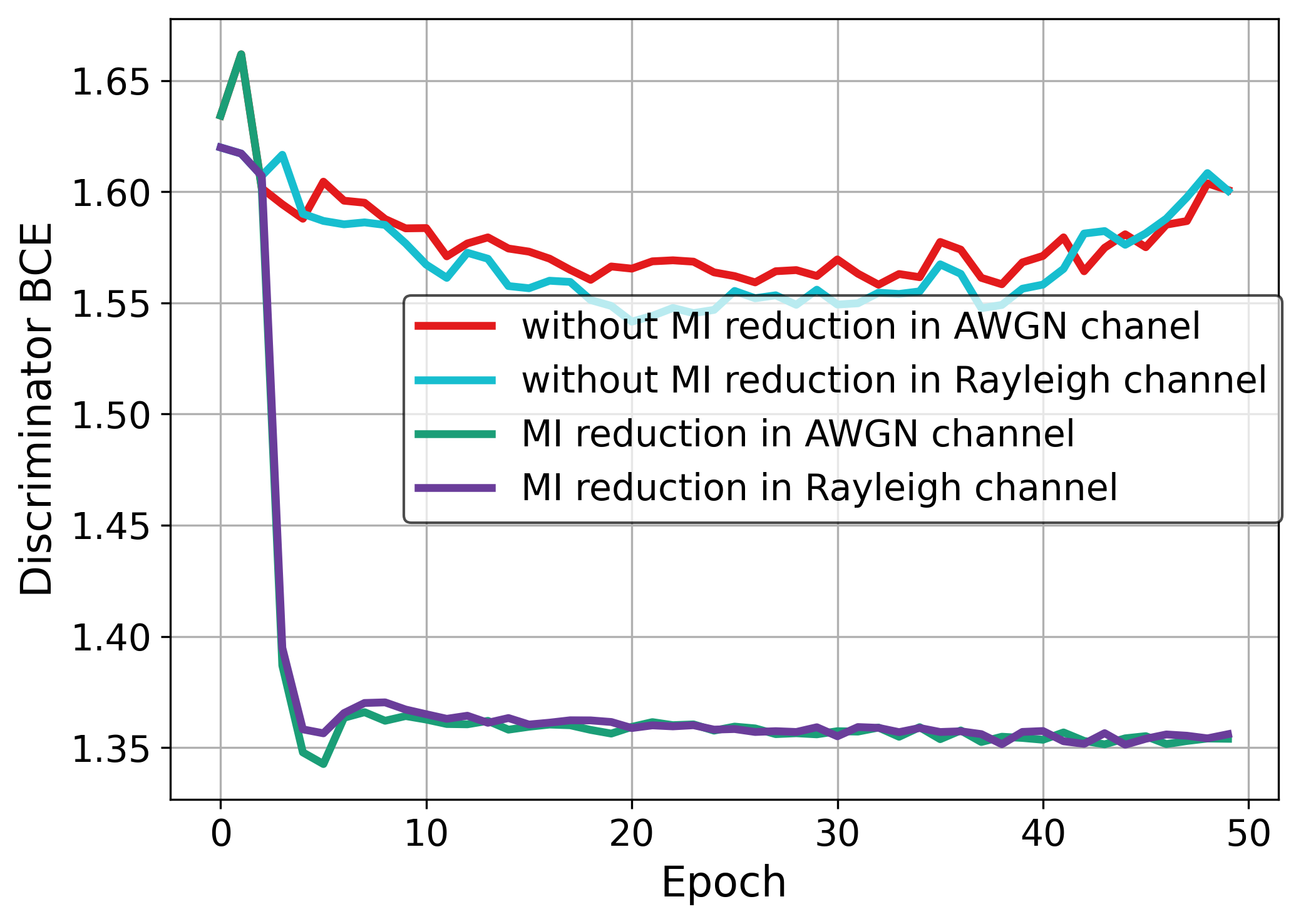}
        \subcaption{BCE between text and image.}
        \label{fig:mi_lv}
    \end{subfigure}%
    \hfill
    \begin{subfigure}[t]{0.32\textwidth}
        \centering
        \includegraphics[width=\textwidth]{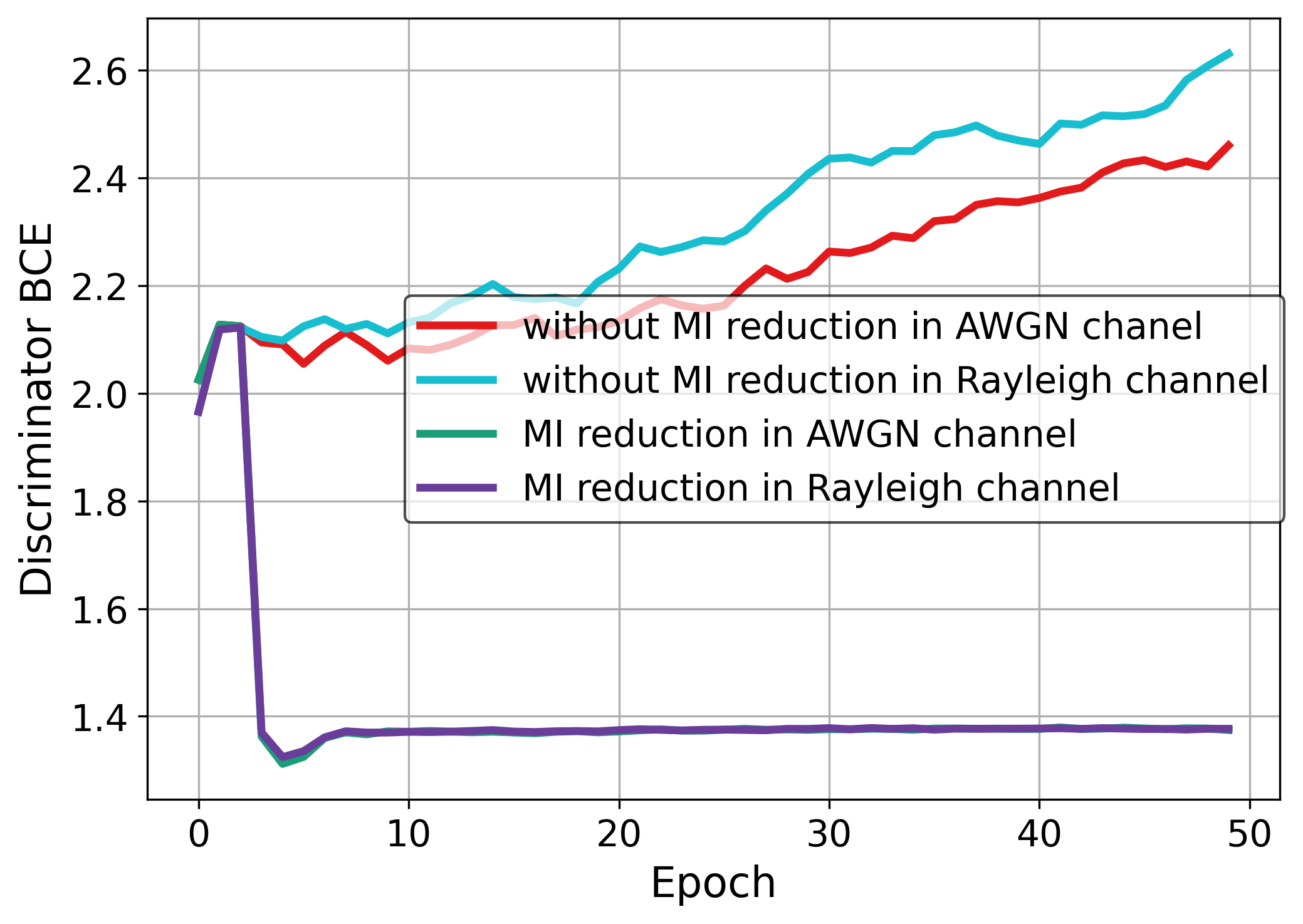}
        \subcaption{BCE between text and audio.}
        \label{fig:mi_la}
    \end{subfigure}
    \caption{Comparison of discriminator BCE between different modalities in MOSEI dataset.}
    \label{fig:mi_comparison_mosei}
\end{figure*}

\begin{figure*}[htbp]
    \centering
    \begin{subfigure}[t]{0.32\textwidth}
        \centering
        \includegraphics[width=\textwidth]{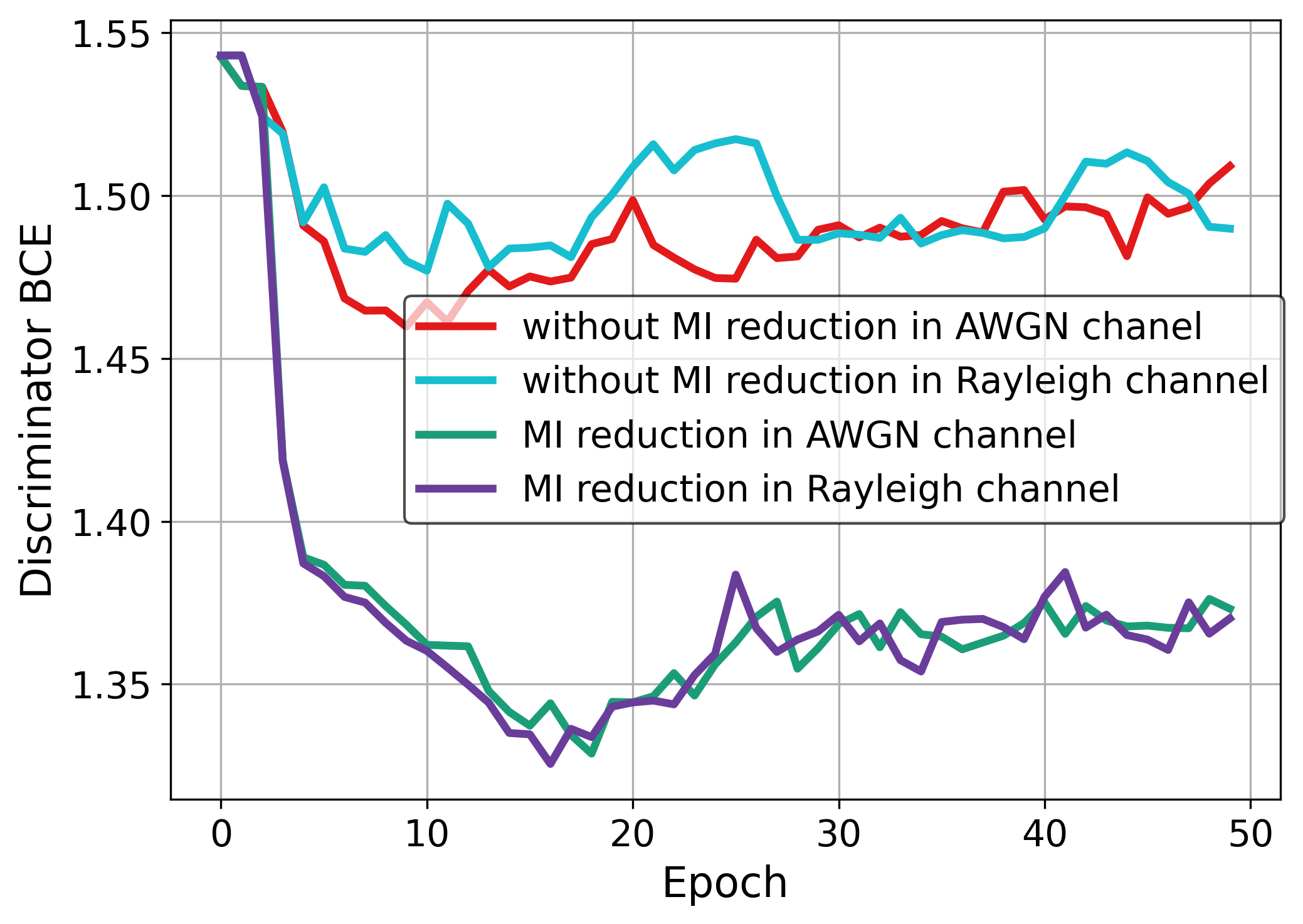}
        \subcaption{BCE between image and audio.}
        \label{fig:mi_va}
    \end{subfigure}%
    \hfill
    \begin{subfigure}[t]{0.32\textwidth}
        \centering
        \includegraphics[width=\textwidth]{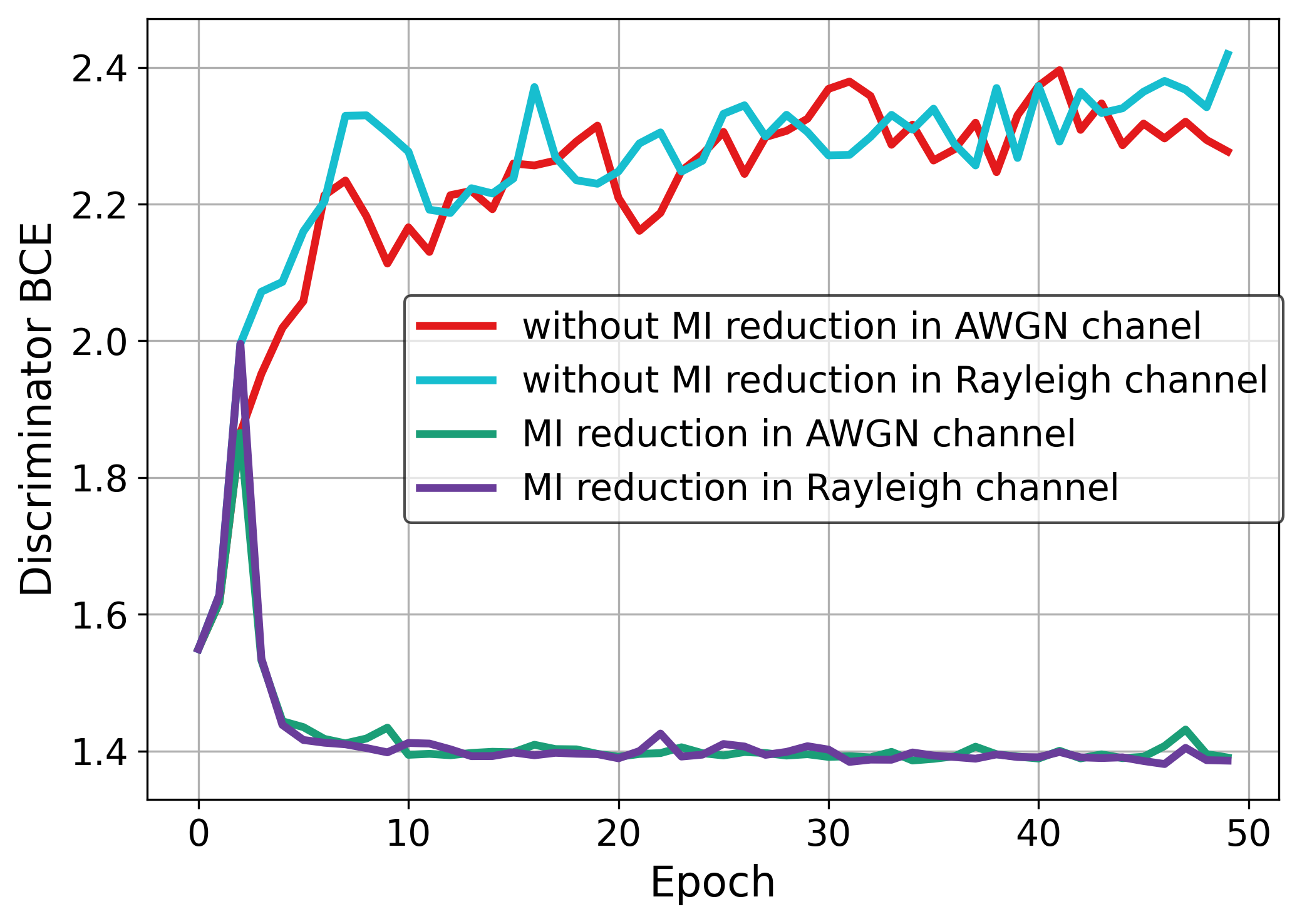}
        \subcaption{BCE between text and image.}
        \label{fig:mi_lv}
    \end{subfigure}%
    \hfill
    \begin{subfigure}[t]{0.32\textwidth}
        \centering
        \includegraphics[width=\textwidth]{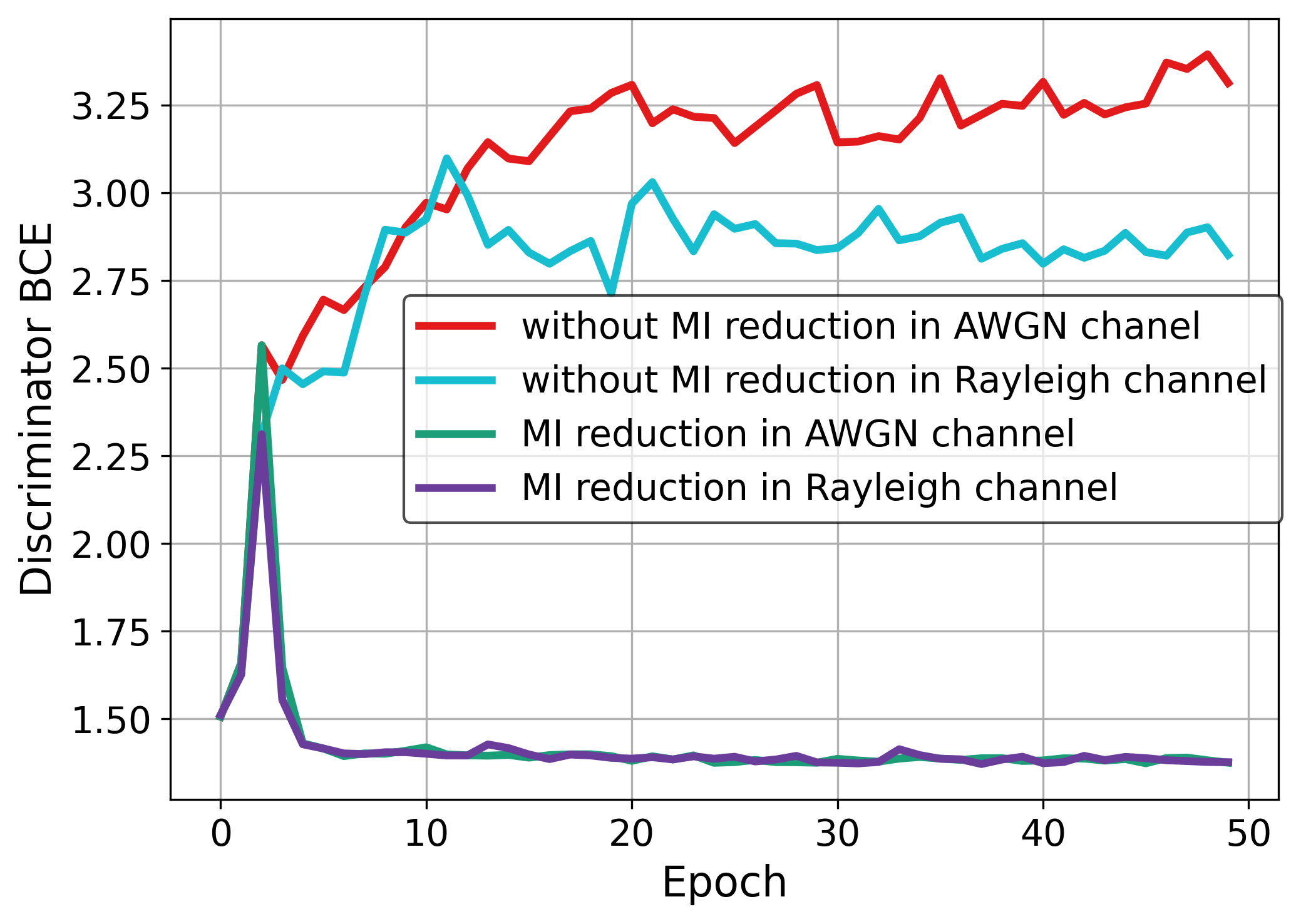}
        \subcaption{BCE between text and audio.}
        \label{fig:mi_la}
    \end{subfigure}
    \caption{Comparison of discriminator BCE between different modalities in MOSI dataset.}
    \label{fig:mi_comparison_mosi}
\end{figure*}

\begin{figure}[t]
    \centering
    \begin{subfigure}{0.7\linewidth}
        \centering
        \includegraphics[width=\linewidth]{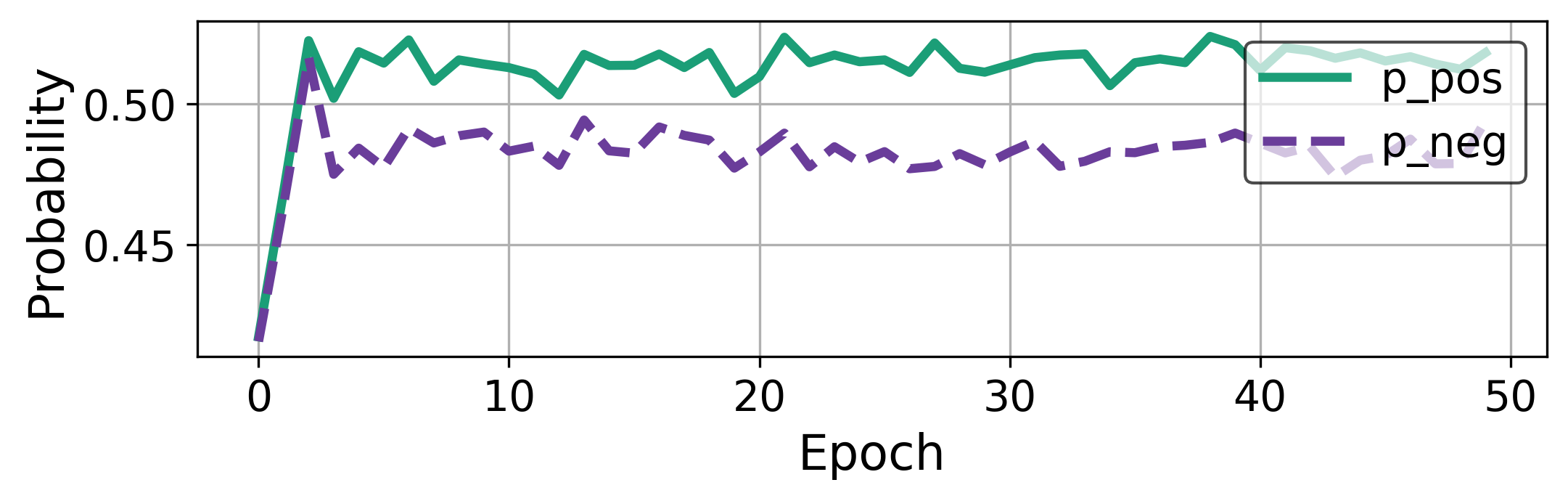}
        \caption{Text and image MI probabilities.}
    \end{subfigure} 
    \begin{subfigure}{0.7\linewidth}
        \centering
        \includegraphics[width=\linewidth]{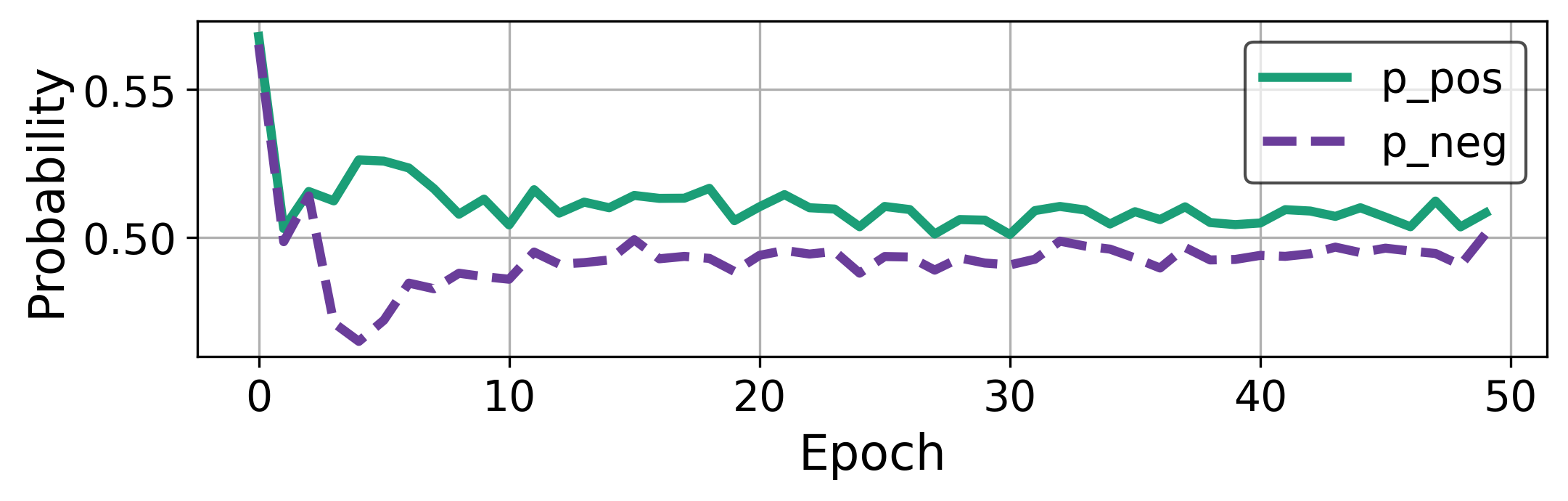}
        \caption{Text and audio MI probabilities.}
    \end{subfigure}
    \begin{subfigure}{0.7\linewidth}
        \centering
        \includegraphics[width=\linewidth]{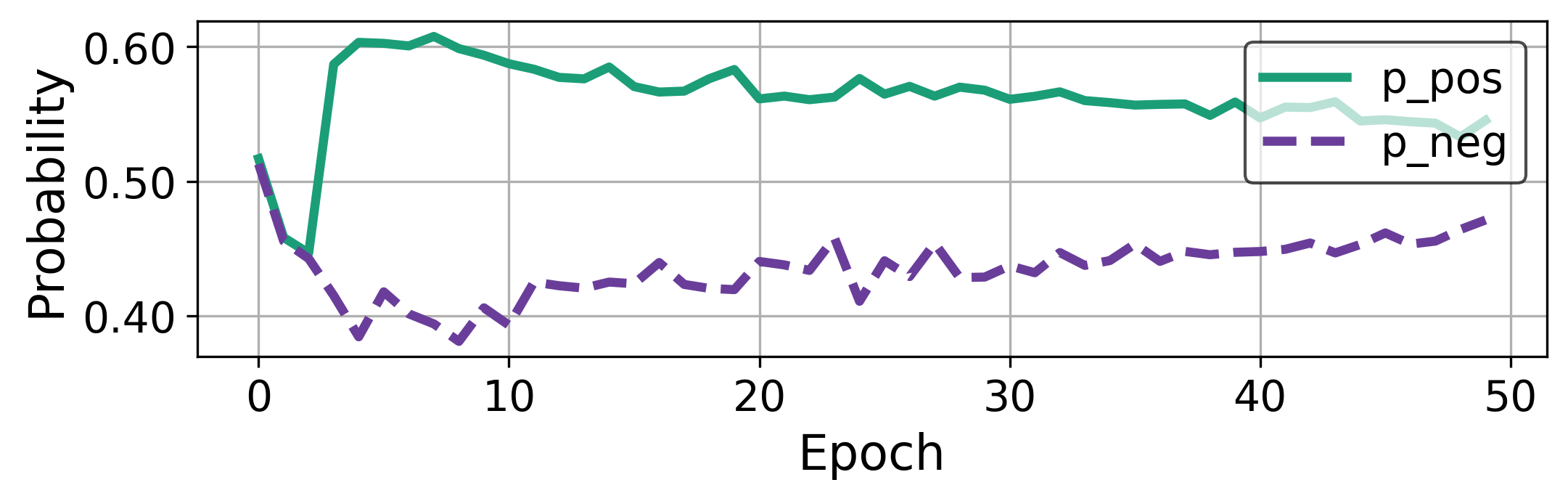}
        \caption{Image and audio MI probabilities.}
    \end{subfigure}
    \caption{MI discriminator probabilities between two modalities.}
    \label{fig:mi_probabilities}
\end{figure}

Figs.~\ref{fig:mi_comparison_mosei} and ~\ref{fig:mi_comparison_mosi} show the discriminator binary cross-entropy (BCE) loss, which we use to evaluate the statistical dependency between modality representations.  BCE loss is inversely related to the JSD-based MI lower bound $J$ (discussed in Sec.~\ref{sec:MI}) via the relation $J \;=\; 2\ln 2 \;-\; \mathrm{BCE}.$
The theoretical maximum for BCE is $2\ln 2 \approx 1.386$ nats, which corresponds to the loss of a random-guessing discriminator ($T(\cdot) \approx 0.5$), which occurs when samples from the joint distribution cannot be distinguished from the product of marginals. Therefore, a BCE value approaching $1.386$ nats directly signifies that $J \approx 0$, indicating that the representations have become statistically independent, and redundant cross-modal information has been successfully removed. As shown in Figs.~\ref{fig:mi_comparison_mosei} and ~\ref{fig:mi_comparison_mosi}, for all modality pairs, the BCE of the discriminator rapidly increases to and stabilizes near the $1.386$ nats baseline. The high, stable loss shows that the discriminator is reduced to random guessing, implying that the learned representations are approximately independent. Occasional transient dips below this level (e.g., on MOSEI dataset in AWGN channels) are expected results of the min-max optimization, where the discriminator momentarily identifies residual dependencies before the GRL updates restore indistinguishability. In contrast, the baseline model without the redundancy reduction consistently exhibits BCE values significantly below $1.386$ nats and/or high volatility. This demonstrates that non-negligible statistical dependence ($J > 0$) persists between the modalities. The robust performance of our method across both MOSEI and MOSI datasets and under both AWGN and Rayleigh channels validates the efficacy of the proposed adversarial redundancy reduction approach.


\subsubsection{Effect of transmitted dimension}

\begin{table}[t]
\centering
\caption{Performance with different transmitted dimensions.}
\label{tab:dl}
\begin{tabular}{c|cccc}
\toprule
\makecell{Transmitted \\ Dimension} & 
Acc2 $\uparrow$ & 
Acc7 $\uparrow$ & 
F1 $\uparrow$   & 
MAE $\downarrow$ \\
\midrule
10 & 0.84227 & 0.48675 & 0.84151 & 0.72271 \\
20 & 0.84724 & 0.52380 & 0.84365 & 0.60373 \\
30 & 0.85110 & 0.50528 & 0.85064 & 0.60918 \\
40 & 0.85110 & 0.51973 & 0.85131 & 0.60090 \\
50 & \textbf{0.85414} & 0.51625 & 0.85430 & \textbf{0.59759} \\
60 & 0.85138 & \textbf{0.52552} & 0.85113 & 0.60183 \\
70 & 0.85352 & 0.51023 & \textbf{0.85531} & 0.60139 \\
80 & 0.85393 & 0.51375 & 0.85493 & 0.60103 \\
\bottomrule
\end{tabular}
\end{table}

To investigate the impact of the latent dimensions of the transmitted signal, we conducted an ablation study. The experiment is performed on MOSEI dataset over AWGN channel with $SNR=12\,dB$. The results for various dimensions are presented in Table~\ref{tab:dl}. As shown in the table, performance generally improves as the transmitted dimension increases from 10 to 50. A significant performance gain is observed when the dimension increases from 10 to 20, particularly in MAE metric. However, the performance gains begin to decrease for more than $50$ dimensions. More dimensions yield only marginal improvements or slight fluctuations in performance, while incurring a higher communication cost. Specifically, a dimension of 50 achieves the best performance for the Top-2 Accuracy and MAE metrics, and its F1 score is highly competitive with the peak performance observed at a higher dimension. Therefore, considering the trade-off between task performance and communication overhead, we selected a transmitted dimension of 50 for all our main experiments, as it offers a compelling balance.

\subsubsection{Validation of MI minimization}

To validate the effectiveness of the adversarial training, we analyze the output probabilities of the cross-modal discriminators, as shown in Fig.~\ref{fig:mi_probabilities}. Each discriminator is tasked with distinguishing paired samples from the joint distribution of two modalities $p_{\text{pos}}$ from unpaired samples drawn from the product of their marginal distributions $p_{\text{neg}}$. The results for all three modality pairs consistently show that both $p_{\text{pos}}$ and $p_{\text{neg}}$ rapidly converge to approximately 0.5 and remain stable. The results indicate that the discriminator has reached an equilibrium where it is unable to distinguish between the two sample types. The outcome serves as strong empirical evidence that the adversarial objective has been met. That is, the uni-modal encoders have successfully learned to generate disentangled representations with minimal statistical dependency. Therefore, the behavior of the discriminator probabilities validates the successful removal of inter-modal redundancy by our proposed framework.



\section{Conclusions}
\label{sec:conclusion}

We propose a two-stage VIB framework to enhance the efficiency and robustness of multi-modal TOC. The framework first applies U-VIB to perform modality-specific compression while retaining task-relevant features. An adversarial cross-modal redundancy reduction method is proposed to further suppress inter-modal dependencies, ensuring that the learned representations are complementary rather than redundant. The second-stage M-VIB further compresses the fused representation, significantly enhancing robustness against channel impairments. Extensive experiments on MOSI and MOSEI datasets validated our design, demonstrating superior performance over baseline models across various channel conditions, particularly in low SNR regimes, by enabling compact and disentangled feature learning. 

\renewcommand*{\bibfont}{}
\printbibliography


\appendices

\section{Derivation of Variational Lower Bounds}
\label{proof:VIB}

In the U-VIB objective
\begin{equation}
    \min_{p(z^m \vert s^m)}\; I(S^m; Z^m) - \beta\, I(Z^m; Y),
\end{equation}
$I(Z^m ; Y)$ can be evaluated as,
\begin{equation}
    I(Z^m ; Y) = \iint dy\, dz^m\, p(y, z^m) \log \frac{p(y|z^m)}{p(y)}.
\end{equation}
 The encoder is $p(z^m|s^m)$, the joint distribution is $p(s^m, y, z^m) = p(s^m,y)p(z^m|s^m)$, we have
\begin{equation}
    I(Z^m ; Y) = \iiint ds^m\, dy\, dz^m\, p(s^m, y) p(z^m|s^m) \log \frac{p(y|z^m)}{p(y)}.
\end{equation}
The posterior $p(y|z^m)$ is often intractable. We introduce a variational distribution $q(y|z^m)$ to approximate it as below,
\begin{equation}
    \log p(y|z^m) = \log q(y|z^m) + \log \frac{p(y|z^m)}{q(y|z^m)}.
\end{equation}
By the non-negativity of the KL divergence, we know  $D_{\mathrm{KL}}(p(y|z^m) \| q(y|z^m)) \ge 0$, which leads to the inequality,
\begin{equation}
    \mathbb{E}_{p(y|z^m)} \big[ \log p(y|z^m) \big] \geq \mathbb{E}_{p(y|z^m)} \big[ \log q(y|z^m) \big].
\end{equation}
Substituting the inequality into the definition of MI, $I(Z^m ; Y) = H(Y) - H(Y|Z^m)$, we have a lower bound
\begin{align}
    I(Z^m ; Y) 
    &= H(Y) - H(Y|Z^m) \notag \\
    &\geq H(Y) + \mathbb{E}_{p(s^m,y,z^m)} \big[ \log q(y|z^m) \big].
\end{align}
Since $H(Y)$ does not depend on the model parameters, it can be treated as a constant during optimization. Thus, we can maximize the lower bound,
\begin{align}
    I(Z^m ; Y)  &\geq \mathbb{E}_{p(s^m,y)p(z^m|s^m)}[\log q(y|z^m)] \notag \\
    &= \iiint ds^m\, dy\, dz^m\, p(s^m,y)\, p(z^m|s^m)\, \log q(y|z^m).
\end{align}
By definition, the MI $I(S^m; Z^m)$ is
\begin{equation}
    I(S^m; Z^m) = \iint ds^m\, dz^m\, p(s^m, z^m) \log \frac{p(z^m|s^m)}{p(z^m)}.
    \label{MI_SMZ}
\end{equation}
The true marginal distribution of the latent variable, $p(z^m) = \int  p(s^m) p(z^m|s^m)ds^m $, is typically intractable because it requires integrating over the entire dataset. To address the problem, we introduce another variational distribution, $q(z^m)$, to approximate $p(z^m)$.
From the non-negativity of the KL divergence, $D_{\mathrm{KL}}(p(z^m) \| q(z^m)) \geq 0$, we have,
\begin{equation}
    \int dz^m\, p(z^m) \log p(z^m) \geq \int dz^m\, p(z^m) \log q(z^m).
\end{equation}
Substituting this into (\ref{MI_SMZ}) gives an upper bound,
\begin{align}
    I(S^m; Z^m) &= \iint ds^m\, dz^m\, p(s^m) p(z^m|s^m) \log \frac{p(z^m|s^m)}{p(z^m)} \notag \\
    &\leq \iint ds^m\, dz^m\, p(s^m) p(z^m|s^m) \log \frac{p(z^m|s^m)}{q(z^m)} \notag \\
    &= \mathbb{E}_{p(s^m)} \Big[ D_{\mathrm{KL}} \big( p(z^m|s^m) \,\|\, q(z^m) \big) \Big]. 
\end{align}
By combining the lower bound for $I(Z^m; Y)$ and the upper bound for $I(S^m; Z^m)$, we can formulate a tractable lower bound for the objective $L_{\text{U-VIB}}$,
\begin{align}
    L_{\text{U-VIB}} 
    &= I(S^m; Z^m) - \beta\, I(Z^m; Y) \notag \\
    &\geq \mathbb{E}_{p(s^m)} \Big[ D_{\mathrm{KL}}\big( p(z^m|s^m) \,\|\, q(z^m) \big) \Big] \notag \\
    &\quad - \beta\, \mathbb{E}_{p(s^m,y)p(z^m|s^m)} \Big[ \log q(y|z^m) \Big].
\end{align}


\section{Proof of Proposition 1}
\label{proof_2log2}

\textit{Step 1: Variational characterization:}
By the f-GAN / JS-divergence variational representation~\cite{nowozin2016f},
\begin{align}
\sup_{T}\Big\{\mathbb{E}_{p}\!\big[\log \sigma(T)\big]
    + \mathbb{E}_{q}\!\big[\log(1-\sigma(T))\big]\Big\} \notag \\
= -2\log 2 \;+\; 2\,D_{\mathrm{JS}}(p\|q).
\end{align}
Hence, at the optimal discriminator \(T^*\),
\begin{align}
\label{eq:J_equals_2JSD}
\mathcal{J}_{\log\sigma}(Z^i;Z^t)
    &= \Big(-2\log 2 + 2\,D_{\mathrm{JS}}(p\|q)\Big) + 2\log 2 \notag \\
    &= 2\,D_{\mathrm{JS}}(p\|q).
\end{align}

\textit{Step 2: Bound and equality conditions for \(D_{\mathrm{JS}}\):}
By definition, we have
\[
D_{\mathrm{JS}}(p\|q)
= \tfrac12 D_{\mathrm{KL}}(p\|m) + \tfrac12 D_{\mathrm{KL}}(q\|m).
\]
\emph{Lower bound \& zero condition.}
Since \(D_{\mathrm{KL}}\ge 0\), we have \(D_{\mathrm{JS}}(p\|q)\ge 0\).
Moreover, \(D_{\mathrm{JS}}(p\|q)=0\) iff \(D_{\mathrm{KL}}(p\|m)=D_{\mathrm{KL}}(q\|m)=0\),
i.e., \(p=m=q\) almost everywhere (a.e.), thus \(p=q\) a.e.

\emph{Upper bound \(\log 2\).}
Because \(m=\tfrac12(p+q)\ge \tfrac12 p\) and \(m\ge \tfrac12 q\),
we have \(\frac{p}{m}\le 2\) on \(\mathrm{supp}(p)\) and \(\frac{q}{m}\le 2\) on \(\mathrm{supp}(q)\).
Therefore,
\begin{align}
D_{\mathrm{KL}}(p\|m) &= \int p \log\frac{p}{m}\,\mathrm{d}x
   \;\le\; \int p \log 2\,\mathrm{d}x = \log 2, \notag \\
D_{\mathrm{KL}}(q\|m) &\le \log 2,
\end{align}
which implies \(D_{\mathrm{JS}}(p\|q)\le \tfrac12(\log 2+\log 2)=\log 2\).
Equality holds iff \(\frac{p}{m}=2\) on \(\mathrm{supp}(p)\) and \(\frac{q}{m}=2\) on \(\mathrm{supp}(q)\) a.e.,
equivalently \(q=0\) on \(\mathrm{supp}(p)\) and \(p=0\) on \(\mathrm{supp}(q)\). That is, \(p\) and \(q\) are
mutually singular (their supports are disjoint a.e.).

\textit{Step 3: Conclusion for \(\mathcal{J}(Z^i;Z^t)\):}
Combining \eqref{eq:J_equals_2JSD} with the bounds above yields
\[
0 \;\le\; \mathcal{J}_{\log\sigma}(Z^i;Z^t)
= 2\,D_{\mathrm{JS}}(p\|q)
\;\le\; 2\log 2.
\]
Moreover, \(\mathcal{J}_{\log\sigma}(Z^i;Z^t)=0\) iff \(D_{\mathrm{JS}}(p\|q)=0\),
i.e., \(p=q\) a.e., which is equivalent to \(p_{z^i z^t}=p_{z^i}p_{z^t}\) and hence
\(I(Z^i;Z^t)=D_{\mathrm{KL}}(p\|q)=0\).
The upper bound \(2\log 2\) is attained in the limit when \(p\) and \(q\) are perfectly separable.

\end{document}